\documentclass[%
 reprint, amsmath, amssymb, aps,]{revtex4-1}

\usepackage{graphicx}
\usepackage{dcolumn}
\usepackage{bm}
\usepackage{color}
\usepackage{url}
\usepackage[normalem]{ulem}
\usepackage{mathtools}
\usepackage{array}
\newcolumntype{P}[1]{>{\centering\arraybackslash}p{#1}}
\newcolumntype{M}[1]{>{\centering\arraybackslash}m{#1}}
\usepackage[caption=false]{subfig}
\usepackage{dcolumn}
\usepackage{graphicx, epsfig}
\usepackage[dvipsnames]{xcolor}
\usepackage{mathrsfs}
\usepackage{yhmath}
\usepackage[caption=false]{subfig}
\usepackage[normalem]{ulem}
\usepackage{mathtools}
\usepackage{bigints}
\usepackage{float}
\usepackage[colorlinks = true, linkcolor = purple, urlcolor  = blue, citecolor = blue, anchorcolor = blue]{hyperref}
\usepackage{float}
\usepackage{multirow}

\newcommand{\RN}[1]{%
  \textup{\uppercase\expandafter{\romannumeral#1}}%
}

\begin{document}

\preprint{APS/123-QED}

\title{Microquasar jet-cocoon systems as PeVatrons}

\author{B. Theodore Zhang$^{1,2}$}
\email[]{zhangbing@ihep.ac.cn}
\author{Shigeo S. Kimura$^{3,4}$}
\email[]{shigeo@astr.tohoku.ac.jp}
\author{Kohta Murase$^{5,6}$}
\email[]{murase@psu.edu}

\affiliation{$^1$ Key Laboratory of Particle Astrophysics and Experimental Physics Division and Computing Center, Institute of High Energy Physics, Chinese Academy of Sciences, 100049 Beijing, China}
\affiliation{$^2$ TIANFU Cosmic Ray Research Center, Chengdu, Sichuan, China}
\affiliation{$^3$ Frontier Research Institute for Interdisciplinary Sciences, Tohoku University, Sendai 980-8578, Japan}
\affiliation{$^4$ Astronomical Institute, Tohoku University, Sendai 980-8578, Japan}
\affiliation{$^5$ Department of Physics; Department of Astronomy \& Astrophysics; Center for Multimessenger Astrophysics, Institute for Gravitation and the Cosmos, The Pennsylvania State
University, University Park, Pennsylvania 16802, USA}
\affiliation{$^6$ Center for Gravitational Physics and Quantum Information, Yukawa Institute for
Theoretical Physics, Kyoto University, Kyoto, Kyoto 606-8502, Japan}

\date{\today}
             
\begin{abstract}
The origin of Galactic cosmic rays (CRs), particularly around the knee region ($\sim$3 PeV), remains a major unanswered question. Recent observations by LHAASO suggest that the knee is shaped mainly by protons, with a transition to heavier elements at higher energies.
Microquasars -- compact jet-emitting sources -- have emerged as possible PeV CR accelerators, especially after detections of ultrahigh-energy gamma rays from these systems. We propose that the observed proton spectrum (hard below a few PeV, steep beyond) arises from the reacceleration of sub-TeV Galactic CRs via shear acceleration in large-scale microquasar jet-cocoon structures.
Our model also naturally explains the observed spectrum of energies around a few tens of PeV by summing up heavier nuclei contributions. Additionally, similar reacceleration processes in radio galaxies can contribute to ultrahigh-energy CRs, bridging Galactic and extragalactic origins. Combined with low-energy CRs from supernova remnants and galaxy clusters around the second knee region, this scenario could provide a unified explanation for CRs across the entire energy spectrum.
\end{abstract}

\pacs{Valid PACS appear here}
\maketitle


\section{\label{sec:1}Introduction}
The origin of cosmic rays (CRs), mainly composed of protons, heliums, and heavier nuclei, is a long-standing puzzle in astrophysics~\cite{KASCADE:2005ynk,Kotera:2011cp, Kachelriess:2019oqu}. 
The energy spectrum of CRs spans over a wide range from $10^9$ to $10^{20}\rm~eV$, roughly following a power-law spectrum $\Phi_{\rm CR}\propto E^{-\gamma}$. 
The spectrum exhibits a distinctive feature with a change in the power-law spectral index from $\gamma \sim 2.7$ to $
\gamma \sim 3$ at the energy around $4\rm~PeV$, known as the kneeknee~\cite{KASCADE:2005ynk, TIBETIII:2008qon, LHAASO:2024knt}, discovered by Kulikov and Khristiansen in 1958~\cite{Kulikov1958}. 
At energies $\sim 100\rm~PeV$, the spectral index changed from $\gamma \sim 3$ to $\gamma \sim 3.3$, called the ``second knee'' \cite[e.g.,][]{Nagano:1991jz, TelescopeArray:2018bya}. 
The spectral index changes back to $\gamma \sim 2.5$ at the ankle around 4 EeV \cite[e.g.,][]{HiRes:1993hwv}, and the flux is largely suppressed at a few tens of EeV~\cite{Abbasi:2007sv, Abraham:2008ru}.
The knee is believed to be formed due to the limited maximum rigidity of Galactic CR accelerators~\cite{2005JPhG...31R..95H}, or reflecting a spectral break at the same rigidity~\cite{Cardillo:2015zda}. The knee could also be formed due to the change of the energy-dependent diffusion coefficient in the propagation of CRs~\cite{Giacinti:2015hva}, while
CRs between the knee and ankle should be marked as the transition region from Galactic to extragalactic CRs~\cite{Gaisser:2006xs}. CRs with energy above EeV are regarded as ultrahigh-energy cosmic rays (UHECRs), which have an extragalactic origin as supported by the observation of the large-scale anisotropy, where the dipole direction points in the opposite direction of the Galactic Center~\cite{PierreAuger:2018zqu}. 

CR composition is the key to identifying the origins.
The general pattern of CR abundances around a few GeV per nucleon resembles that of the interstellar medium or primordial solar composition, with peaks in the carbon-nitrogen-oxygen (CNO) and Fe region~\cite{Drury:2018mkl}. 
Recently, a kneelike feature around $3\rm~PeV$ was also revealed on the measured all-particle spectrum of CRs. 
The logarithm of the mean mass of CRs measured by the Large High Altitude Air Shower Observatory (LHAASO) showed trends toward a heavier composition below several hundred TeV and above several PeV, while it is a light composition dominated around the knee, and the turning point is consistent with the position of the knee at energy around $3\rm~PeV$~\cite {LHAASO:2024knt}. 
In particular, LHAASO reported the first high-purity proton spectrum, implicating the knee at energy around $3\rm~PeV$ is related to the change of the proton spectrum~\cite{LHAASO:2025byy}. The observed hardening feature above several hundreds of TeV implicates the emergence of a new component that has maximum energy at least up to a few PeV~\cite{LHAASO:2025byy}.

The physical origin of the knee is still unknown and should be related to Galactic PeVatrons. 
Supernova remnants (SNRs)  are believed to be responsible for CRs at least up to the knee~\cite{2008ARA&A..46...89R, Blasi:2013rva}. However, SNRs as PeVatrons are challenged by both theoretical and observational studies \cite{2004MNRAS.353..550B, 2005MNRAS.358..181B,  Gabici:2019jvz, Giuliani:2024fpq}, potentially implying the existence of another type of PeVatrons. 
Potential candidates for PeVatron include massive stars and stellar clusters~\cite{1982ApJ...258..860C, 1983SSRv...36..173C, Aharonian:2018oau}, pulsars and pulsar winds \cite{Bednarek:2001av, 2002JPhG...28.1275G, Bednarek:2004wp, Ohira:2017bxa}, superbubbles \cite{2020SSRv..216...42B}, hypernovae~\cite{Sveshnikova2003} and interacting supernovae \cite{Murase:2013kda,Inoue:2021bjx}, and isolated black holes~\cite{Ioka:2016bil, Kimura:2024izc}. 
The detection of very-high-energy (VHE) gamma rays has shown that microquasars are promising sites for accelerating CRs around PeV and above, indicating potential sources of Galactic PeV CRs~\cite{LHAASO:2024psv}. 
Microquasars are x-ray binaries with relativistic jets~\cite{Mirabel:1994rb}
and display phenomenology that is similar to active galactic nuclei (AGNs), although the size of the system is orders of magnitude different. 
It is suggested that shear and one-shot acceleration in the large-scale jet of radio galaxies can be responsible for the origin of UHECRs~\cite{Kimura:2017ubz, Caprioli:2015zka, Mbarek:2019glq, 
Mbarek:2022nat,
Mbarek:2024nvv}.
This motivates us to investigate whether the shear acceleration by microquasar jets can be Galactic PeVatrons. 
Previous studies also suggest microquasars as PeVatrons \cite{Cooper:2020tzq,Kimura:2021eik,Yue:2024nns}, but these studies consider CR acceleration by shocks in jets or magnetic reconnections in accretion flows. 

The paper is organized as follows.
In Sec.~\ref{sec:phy}, we study the physics of Galactic microquasars and the shear acceleration process in the jet-cocoon system of microquasars.
In Sec.~\ref{sec:obs}, we propagate the escaped CRs from microquasars and compare the results with observations. 
In Sec.~\ref{sec:connection}, we combine CRs from Galactic SNRs and contributions from the shear acceleration process from extragalactic clusters of galaxies and radio galaxies, and we proposed an overall picture of the origin of CRs on a wide energy range from GeV to $\sim 100\rm~EeV$.
We give a summary and discussion in Sec.~\ref{sec:sum}.

\section{Cosmic rays accelerated in microquasars}\label{sec:phy}

\subsection{The physics of microquasar jets}\label{sec:microquasar}
\begin{figure}
    \centering
\includegraphics[width=1.0\linewidth]{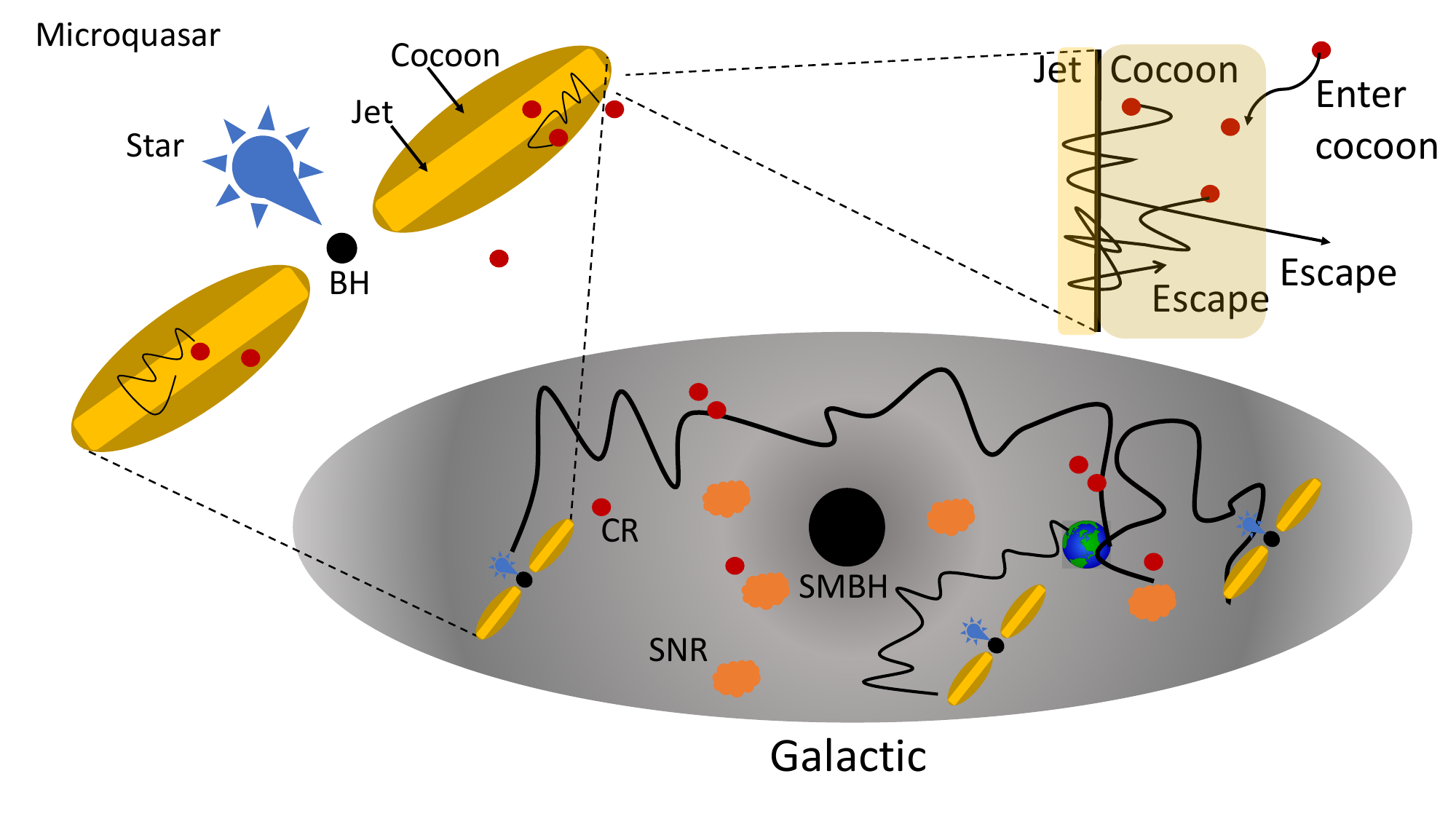}
    \caption{A schematic plot of the reacceleration of Galactic low-energy GeV-TeV CRs in the large-scale jet-cocoon system of microquasars.}
    \label{fig:mq}
\end{figure}

A microquasar consists of a compact object, usually a black hole, and a companion star, see Fig.~\ref{fig:mq} for a schematic plot. 
Currently, over 20 microquasars have been identified in our Galaxy \cite{tetarenko:2016aa,Corral-Santana:2016aa}.
In the microquasar systems, material of the companion star falls to the compact star, which releases a large amount of the gravitational energy. A part of the gravitational energy is converted to outflow kinetic energy. These outflows, called jets, are launched at the vicinity of the compact star, narrowly collimated to a small solid angle, and often reach relativistic speeds. The kinetic power of the jets can reach the Eddington luminosity, $L_j\sim L_{\rm Edd}\sim1\times10^{39}(M/10 M_\odot)\rm~erg~s^{-1}$. The jets propagate in interstellar medium and extend to $\sim100$ pc scales, which is many orders of magnitude larger than the scale of the compact star. See Ref. \cite{Done2007aa} for a comprehensive review.

Recently, Ref.~\cite{LHAASO:2024psv} reported detection of gamma rays above 100 TeV from five microquasars, three of which (SS 433, V4641 Sgr, and GRS 1915+105) exhibited extended gamma-ray morphology whose emission zone is as large as $\sim50-100$ pc . These gamma- ays are most likely originating from the radiation of nonthermal particles accelerated in the extended jets \cite{HAWC:2017osk, HESS:2024rlh, LHAASO:2024psv}.
The other two (Cyg X-1 and MAXI J1820+070) are consistent with point source, possibly suggesting a compact emission region \cite{Fang:2024wmf,Kuze:2025wda}, although the best fit positions of gamma-ray sources have slight offsets from those of x-ray binaries.

SS 433 is a binary system consisting of a compact object and a supergiant star, with a pair of jets moving in opposite directions with launch velocity $v\sim 0.26\rm~c$, where $c$ is the speed of light \cite{Abell:1979aa}. 
The extended two-sided jets with $\sim40$ pc are detected in x-rays \cite{Safi-harv:1997aa}, and radio lobes surrounding the system extend up to $\sim150$ pc \cite{Dubner:1998aa,Sakemi:2023aa}. 
VHE gamma-ray emission from SS 433 has been detected at the jet lobes, which are spatially coincident with the X-ray hot spots~\cite{HAWC:2018gwz, HESS:2024rlh, LHAASO:2024psv, HAWC:2024ysp}.
The observed gamma-ray emission is consistent with inverse Compton emission from a single population of electrons, and the magnetic field strength inside the jet is $\sim 20\rm~\mu G$~\footnote{In Table~1 of Ref.~\cite{Kimura:2020aa}, $B$ is given in the unit of $\mu$G not G.}. 
References~\cite{Sudoh:2020aa,Kimura:2020aa} pointed out that nonthermal electrons accelerated at SS 433 jets achieve PeV energies with a small Bohm factor, based on multiwavelength modeling. 
In addition, Ref.~\cite{Ohmura:2021skq} performed magnetohydrodynamics (MHD) simulations of the propagation of the jet in the surrounding medium, where the jet is spheroidal at early times and the wings are developed by the broadening spherical cocoon.

V4641 Sgr is another interesting microquasar that emits gamma-rays above 100~TeV. The morphology of the gamma-ray signal is clear two-sided lobe structure, and the size of the lobe is $\sim100$ pc~\cite{LHAASO:2024psv, Alfaro:2024cjd}. A spatially extended x-ray signature is also detected from this object, although the x-ray signal is concentrated on the central region within $\sim15$~pc from the central black hole \citep{Suzuki:2024rzc}. The multiwavelength modeling of this system indicates $B\gtrsim 10 \rm\mu G$ at the x-ray emission region.

GRS 1915+105 is a highly variable object that exhibits superluminal motion with extended jets of sub-pc scales \cite{Fender:1998nq}. Its Eddington ratio used to be $\sim0.1-1$ \citep{Done:2003sz} and suddenly decreased to 0.01 in 2018 \cite{Koljonen:2021dqn}. The gamma-ray source detected by LHAASO exhibits an extension of $\sim50$~pc \cite{LHAASO:2024psv}, which is consistent with the two-sided lobe structure detected by ALMA \cite{Tetarenko:2017zke}.

The microquasar S26 in the Sculptor Galaxy NGC 7793 consists of a black hole and a Wolf-Rayet star. Its nebula spans $350 \times 185\rm~pc$~\cite{Dopita:2012aa}, and the jet’s kinetic luminosity $L_j \sim 5 \times 10^{40}\rm~erg~s^{-1}$ suggests super-Eddington accretion, making it a potential PeVatron ~\cite{Abaroa:2024vnw}.

Nonthermal particles can be accelerated through the energy dissipation within the jet. 
With the luminosity requirement~\cite{Blandford:1999hi} obtained from the Hillas condition~\citep{Hillas:1985is}, the maximum possible energy of the accelerated CRs is estimated to be
\begin{align}
E_{A, \rm max}&\approx 3\times{10}^{16}~{\rm eV}~Z \epsilon_B^{1/2}  \nonumber \\
&\times  {\left(\frac{L_j}{{10}^{39}~{\rm erg/s}}\right)}^{1/2} \left(\frac{\Gamma}{1.05}\right)^{-1} \left(\frac{\beta}{0.3}\right)^{1/2}
\end{align}
where $Z$ is the nuclear charge, $\epsilon_B$ is the energy fraction carried by the magnetic field, $L_j$ is the total isotropic-equivalent luminosity of the jet, $\Gamma$ is the Lorentz factor, and $\beta$ is the characteristic velocity normalized by the speed of light.
In our Galaxy, there are likely a few dozen black hole (BH) X-ray binaries with an x-ray luminosity of $L_X \gtrsim 10^{38}\rm~erg~s^{-1}$, each exhibiting sufficient luminosity, $L_j\sim10^{39}\rm~erg~s^{-1}$, to potentially accelerate CRs up to the knee energy range.
Here, we use frequently used assumptions of $L_X\sim0.1\dot{M}c^2$ and $L_j\sim\dot{M}c^2$ to convert $L_X$ to $L_j$, where $\dot{M}$ is the mass accretion rate onto the BH.

\subsection{Shear acceleration in microquasar jets}
Shear acceleration, a class of the Fermi acceleration, has been suggested to be an efficient mechanism to accelerate CRs, where charged particles gain energy inside an ordered shear velocity field~\cite{1981SvAL....7..352B, 1988ApJ...331L..91E, 1990A&A...238..435O, 1989ApJ...340.1112W, Rieger2004, Rieger:2019uyp, 2021MNRAS.503.5948M, Wan:2025eoh}. 
It has been shown that the discrete shear acceleration~\cite{1990A&A...238..435O} occurring on the surface of the jet-cocoon system in radio galaxies may accelerate low-energy ``seed'' galactic CRs to the UHE energy range, and the predicted spectrum and composition are consistent with observations~\cite{Kimura:2017ubz, Seo:2023dkc}.

In AGNs, the reacceleration may occur through the ``one-shot'' acceleration, where the seed CRs receive a boost of a factor of $\Gamma^2$~\cite{Gallant:1998uq}, and it is possible for powerful radio galaxies with a Lorentz factor of $\Gamma \gtrsim 30$ to accelerate seed galactic CRs with energy $E \lesssim 10^{17}\rm~eV$ to the UHE energy range~\cite{Caprioli:2015zka,Mbarek:2019glq}. 
However, the one-shot acceleration is challenging in microquasars because of their low Lorentz factor jets, typically $\Gamma\lesssim3$.

We consider the shear acceleration by cylindrical jets and cocoon as discussed in Kimura, Murase and Zhang (hereafter KMZ18)~\cite{Kimura:2017ubz}. Based on our discussion in Sec.~\ref{sec:microquasar}, we assume that the radius of the jet is $r_{\rm jet} = 3 \rm~pc$, the length is $l_{\rm jet} = 100\rm~pc$, the jet velocity normalized by the speed of light is $\beta_j=0.3$ (corresponding to Lorentz factor $\Gamma_j\simeq1.05$), and the radius of the cocoon is $r_{\rm coc} = 60\rm~pc$.
The coherence length inside the cocoon is set to $l_{\rm coh} \sim 0.03 r_{\rm coc} \sim 1.8\rm~pc$.
The mean free path of CRs is given by $\lambda \sim l_{\rm coh} (E/E_{\rm coh})^\delta$, where $\delta = 1/3$ for Kolmogorov turbulence and $E_{\rm coh} = Z e B_{\rm coc} l_{\rm coh} \sim 
17Z(B_{\rm coc}/10{\rm~\mu G}) (l_{\rm coh}/1.8\rm~pc)\rm~PeV$. 
CRs will undergo a discrete shear acceleration when the CR mean free path is longer than the width of the sheath layer, $\lambda \gtrsim r_{\rm sl}$. For $r_{\rm sl} \sim f_{\rm sl} r_{\rm jet} \sim 0.06 (f_{\rm sl}/0.02)\rm~pc$, the injection energy is given by $\lambda \sim r_{\rm sl}$, which can be rewritten as
\begin{eqnarray}
E_{\rm inj} &\approx& E_{\rm coh} \left(\frac{r_{\rm sl}}{l_{\rm coh}}\right)^3\label{eq:Einj}\\
&\sim& 0.6\left(\frac{B_{\rm coc}}{\rm 10~\mu G}\right)\left(\frac{r_{\rm sl}}{0.06\rm~pc}\right)^3\left(\frac{l_{\rm coh}}{1.8\rm~ pc}\right)^{-2}\rm~TeV.\nonumber
\end{eqnarray}
Thus, CRs with energies above several hundreds of GeV can undergo a discrete shear acceleration mechanism. 
In this context, injection energy is defined as the minimum kinetic energy required for a CR particle to undergo discrete shear acceleration.
We should note that the width of the shear layer and injection energy to the discrete shear acceleration are still uncertain, and $f_{\rm sl}\sim1$ is suggested by recent high-resolution MHD simulations when the kink instability is fully developed \cite{Wang:2024ijr}.

Based on KMZ18~\cite{Kimura:2017ubz}, by balancing the acceleration and effective confinement timescales, the peak energy in the energy spectrum of escaping CR protons can be estimated to be
\begin{eqnarray}
    E_{\rm max} &=& \zeta e B_{\rm coc} l_{\rm coh}^{1/2} r_{\rm jet}^{1/2} \Gamma_{\rm jet} \beta_{\rm jet} \\ 
    &\simeq& 4.3{\rm~PeV} \left(\frac{B_{\rm coc}}{10\mu \rm~G}\right) \left(\frac{l_{\rm coh}}{1.8\rm~pc}\right)^{1/2} \left(\frac{r_{\rm jet}}{3\rm~pc}\right)^{1/2} \left(\frac{\beta_{\rm jet}}{0.3}\right)\zeta,\nonumber
\end{eqnarray}
where $\zeta \equiv (\zeta_c / \zeta_a)^2$ is a numerical factor of order unity. 

The average rate of CRs injected into the shear acceleration is estimated to be
\begin{align}
    \dot{N}_{\rm CR} &\approx \frac{N_{i, \rm inj}}{t_{\rm ad}}  \approx \frac{2\pi r_{\rm jet}^2 l_{\rm jet} n_{\rm cr}}{r_{\rm coc} / v_{\rm exp}} \nonumber \\ 
    &\simeq 7.1 \times 10^{33}{\rm~s^{-1} }\left(\frac{r_{\rm jet}}{3\rm~pc}\right)^2 \left(\frac{l_{\rm jet}}{100\rm~pc}\right) \\ &\times \left(\frac{r_{\rm coc}}{60\rm~pc}\right)^{-1} \left(\frac{v_{\rm exp}}{3000\rm~km~s^{-1}}\right),\nonumber
\end{align}
where $t_{\rm ad} \approx r_{\rm coc} / v_{\rm exp}$ is the adiabatic cooling time, $v_{\rm exp}$ is the expansion velocity and $n_{\rm cr} \sim 2.2 \times 10^{-14}\rm~cm^{-3}$ is the observed number density of CRs at rigidity 0.32~TV. 
Assuming that CRs are accelerated to PeV energies, the total luminosity of CRs accelerated by a microquasar is roughly estimated to be 
\begin{align}
L_{\rm CR} &\approx E_{\rm max}\dot{N}_{\rm CR} \label{eq:Lcr_shear}\\
&\simeq 5.0 \times 10^{37}\left(\frac{r_{\rm jet}}{3\rm~pc}\right)^{5/2} \left(\frac{l_{\rm jet}}{100\rm~pc}\right)\left(\frac{v_{\rm exp}}{3000\rm~km~s^{-1}}\right) \nonumber \\ 
&\times \left(\frac{r_{\rm coc}}{60\rm~pc}\right)^{-1} \left(\frac{B_{\rm coc}}{10\mu \rm~G}\right) \left(\frac{l_{\rm coh}}{1.8\rm~pc}\right)^{1/2} \left(\frac{\beta_{\rm jet}}{0.3}\right)\zeta\rm~erg~s^{-1}.\nonumber
\end{align}

\begin{table}[]
    \caption{Model parameters used in Monte Carlo simulations. }
    \centering
    \begin{tabular}{l|ccccccc}
       \hline
    Parameters & $r_{\rm jet}$ & $B_{\rm jet}$ & $\beta_{\rm jet}$ & $r_{\rm coc}$ & $B_{\rm coc}$ & $l_{\rm coh}$ & $l_{\rm jet}$  \\
     & [pc] & [$\rm \mu G$] &  & [pc] & [$\rm \mu G$] & [pc] & [pc] \\
       \hline
     Value & 3 & 50 & 0.3 & 60 & 10  & 1.8 & 100 \\
    \end{tabular}
    \label{tab:parameters-default}
\end{table}

\begin{figure}
    \centering
\includegraphics[width=1.0\linewidth]{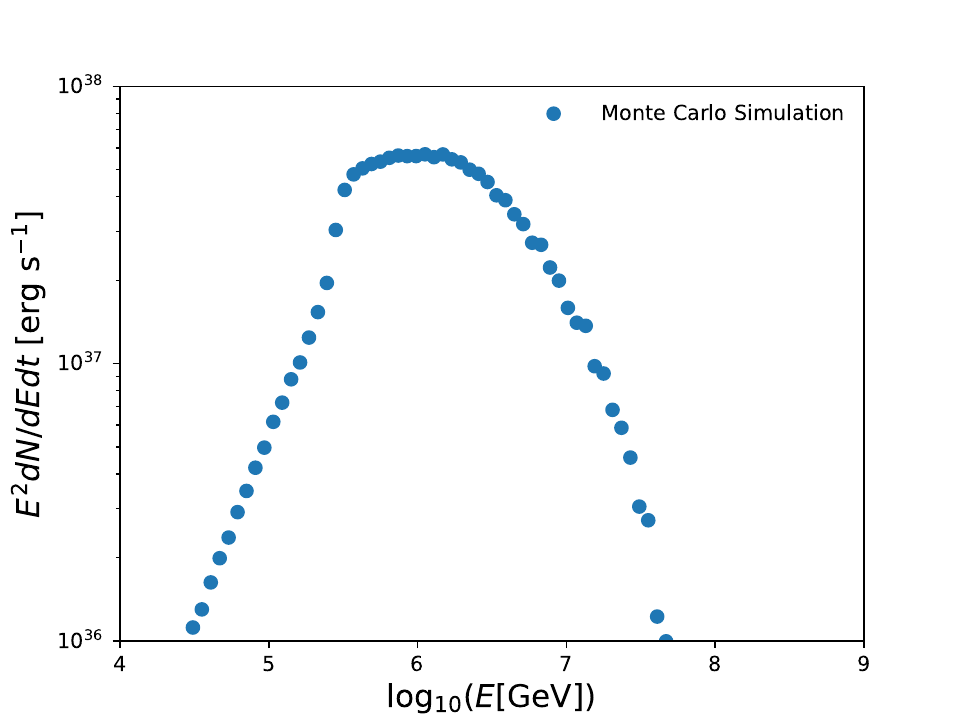}
    \caption{The energy spectrum of escaping CRs predicted in the shear acceleration mechanism from Monte Carlo simulations. The parameters are listed in Table~\ref{tab:parameters-default}. The above plot is multiplied with $\rm exp(-\mathcal{R}/\mathcal{R}_{\rm max})$, where $\mathcal{R}_{\rm max} = 4\times 10^7\rm~GeV$ is the maximum acceleration energy from the microquasar jets. 
    }
    \label{fig:spectrum-inj}
\end{figure}

We perform Monte Carlo simulations with the code used by KMZ18~\cite{Kimura:2017ubz} with parameters given in Table~\ref{tab:parameters-default}, and the resulting energy spectra {and luminosity of the} escaping CR protons are shown in Fig.~\ref{fig:spectrum-inj}. As seen in the figure, our shear acceleration scenario can achieve spectral peak around PeV energies with reasonable parameters (see Sec.~\ref{sec:sum} for a discussion about parameter choices). 
The escaping CR proton spectrum shows a hard component (spectral index $\sim$ 0.8) below the peak energy and
a gradual cutoff above it.

\section{Observed cosmic-ray spectrum on Earth}\label{sec:obs}

\begin{figure}
    \centering
    \includegraphics[width=1.\linewidth]{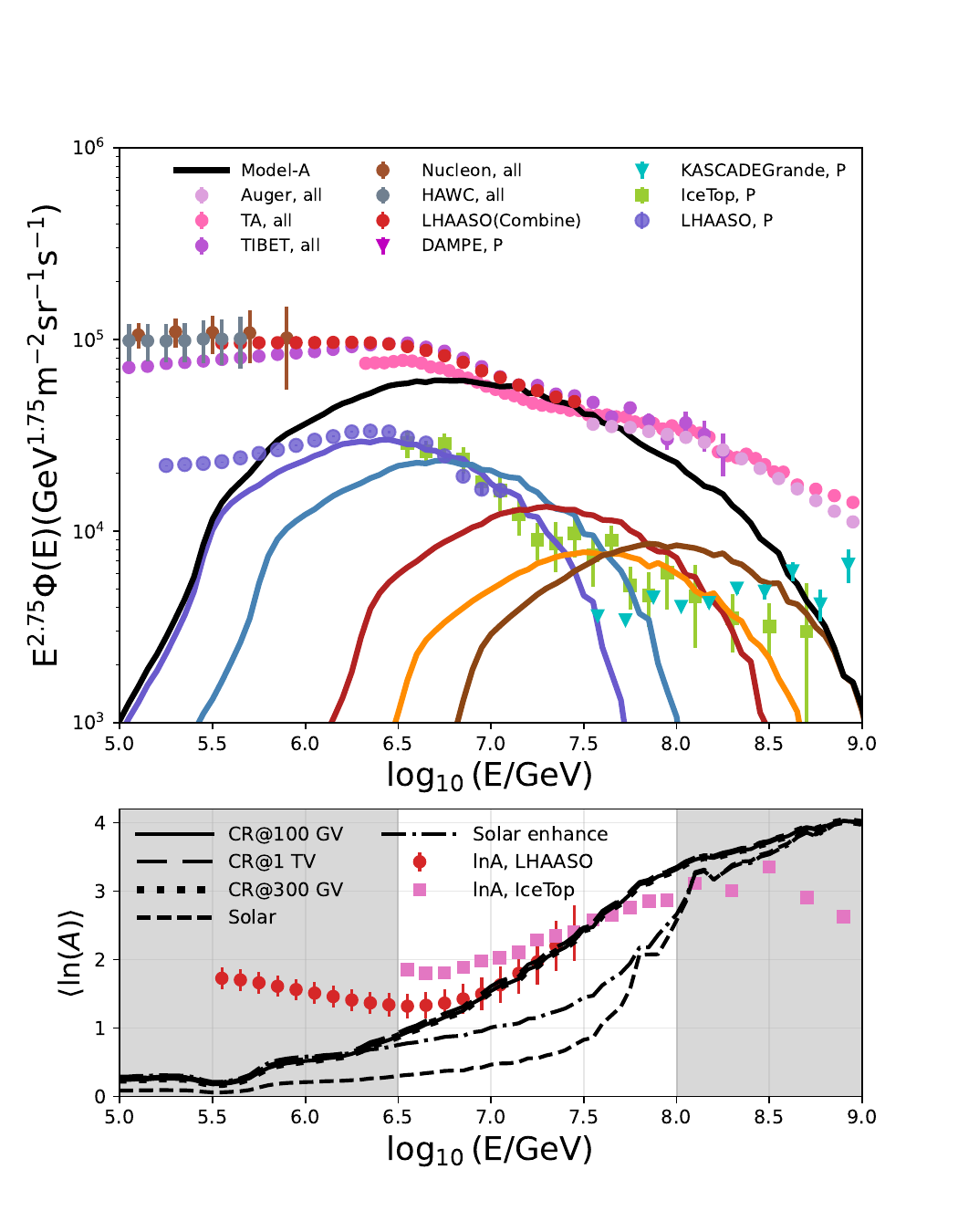}
    \caption{Predicted CR intensity on Earth and the mean mass $\langle {\rm ln}(A) \rangle$ distribution from microquasars.
    The all-particle energy spectra are taken from LHAASO~\cite{LHAASO:2024knt}, HAWC~\cite{HAWC:2017osk}, NUCLEON~\cite{2019AdSpR..64.2546G}, TIBET~\cite{TIBETIII:2008qon}, TA~\cite{Matthews:2017rW}, and Auger~\cite{Verzi:2019AO}, while the proton spectra are taken from LHAASO~\cite{LHAASO:2025byy}, KASCADE-Grande~\cite{Arteaga-Velázquez:2017/0}, and IceTop~\cite{IceCube:2019hmk}.
    The measurements of the mean mass $\langle {\rm ln}(A) \rangle$ distribution are taken from LHAASO~\cite{LHAASO:2024knt} and IceTop~\cite{IceCube:2019hmk}. 
    }
    \label{fig:knee-mq}
\end{figure}

In general, the propagation of CRs in our Galaxy can be described by the transport equation~\cite{1965P&SS...13....9P, Gaisser:2016uoy},
\begin{align}
    &\frac{\partial N_i(E, \mathbf{x})}{\partial t} + \mathbf{V} \cdot \nabla N_i(E, \mathbf{x}) - \nabla \cdot \left[D(r) \nabla N_i(E, \mathbf{x})\right] \nonumber \\
&= Q_i(E, \mathbf{x}, t) - p_i N_i(E, \mathbf{x}) \nonumber \\ &+ \frac{v \rho(\mathbf{x})}{m_p} \sum_{k \geq i} \int \frac{d\sigma_{i,k}(E, E')}{dE} N_k(E', \mathbf{x}) dE',
\end{align} 
where $N_i(E, \mathbf{x})$ is the number density per energy at position $\mathbf{x}$, and the second and third terms on the left side describe the time evolution of the particle number density due to advection and spatial diffusion, respectively. The first term on the right side represents the injection of particles of type $i$, and the second and last terms account for the loss and gain processes. 
Here, the loss of nuclei of type $i$ caused by collisions and decay with a rate $p_i = v\rho/\lambda_i + 1/\gamma \tau_i$, where $\rho$ is the interstellar medium density, $\lambda_i$ is the mean free path of interactions when particle propagates inside the Galaxy, $\tau_i$ is the decay time of a particle of type $i$, and $\gamma$ is theparticle Lorentz factor.
In this study, we employ a publicly available version of the \textsc{DRAGON} code to propagate CRs from sources to Earth~\cite{Evoli:2016xgn}. 
The maximum value of the galactocentric radius is $R_G = 10 \rm~kpc$ and the halo size is $H_G = 4\rm~kpc$.
The spatial distribution of microquasars is unknown.
High-Eddington ratio microquasars, including SS 433 or Sgr V4641, likely follow the distribution of massive stars, and thus, they have similar spatial distribution with Galactic SNRs. 
For the source distribution, we adopt the Ferriere2001 model~\cite{Ferriere:2001rg}, based on the population studies of SNRs.
The spatial distribution of the diffusion coefficient is constant and parametrized as $D = 23.8 \times 10^{28} (\mathcal{R}/250\rm~GV)^\delta {\rm cm}^{2} {\rm~s}^{-1}$ with $\delta = 0.45$ for $\mathcal{R} < 250\rm~GV$, while $\delta = 1/3$ for $\mathcal{R}\geqslant 250\rm~GV$~\cite{Murase:2018utn, Kimura:2018ggg}.
The chosen diffusion coefficient is determined by the observed boron to carbon (B/C) ratio~\cite{AMS:2016brs}. 

We adjust the energy injection rate of CRs to match the CR data at PeV energies, which requires $Q_{\rm cr, tot} \sim 1.3 \times 10^{42}\rm~erg~kpc^{-3}~yr^{-1}$. The energy injection rate integrated over the Galactic volume is 
\begin{equation}
    L_{\rm cr, tot} = Q_{\rm CR, tot} V_G \simeq 1.5\times 10^{38}\rm~erg~s^{-1},\label{eq:Lcr_tot}
\end{equation}
where $V_G \sim 2 \pi R_G^2 \times H_G  \sim 1 \times 10^{68}\rm~cm^3$ is the Galactic volume. 
Our scenario is consistent with the value required to explain CRs in the knee energy range if we have around three to ten microquasars in our Galaxy. 
For our scenario to work, microquasars need to have extended jets of $\sim100$~pc. Only two objects (SS433 and V4641 Sgr) exhibit such a feature. 
The other three microquasars, GRS 1915, Cyg X-1, MAXI J1820, do not have such large-scale extended jets, although they have more compact jets. Since we can observe only less than half of our Galaxy, three to ten microquasars with extended jets sounds more reasonable.
Note that the required number of microquasars with extended jets depends on the injection energy given by Eq.~\ref{eq:Einj}.

In Fig.~\ref{fig:knee-mq}, we present the predicted energy spectrum and mean mass $\langle {\rm ln}(A) \rangle$ of CRs from Galactic microquasars in our shear acceleration scenario. 
Note the shaded region in the lower panel about $\langle {\rm ln}(A) \rangle$ are affected by low- and high-energy components.
In our scenario, the CR composition ratio is determined by the Galactic CR composition ratio at the rigidity of $\mathcal{R}_{\rm inj}=E_{\rm inj}/Z$. Since $E_{\rm inj}$ is uncertain, we examine several values of $E_{\rm inj}$ (see the next paragraph). In Fig.~\ref{fig:knee-mq}, we adopt the CR composition at 300 GV from the AMS-02 observation ~\cite{AMS:2018qxq, AMS:2018cen, AMS:2015tnn, AMS:2021lxc, AMS:2020cai}.
Our model demonstrates that CRs from microquasars could explain both proton and all-particle spectrum around the knee.
The hard proton spectrum below the peak energy (at a few PeV) and a steep power-law decay beyond the peak predicted by our model agrees well with the LHAASO observations. Furthermore, the model self-consistently predicts the spectra of helium, intermediate mass, and heavy nuclei around knee, with their relative abundance normalized to the local Galactic CR abundance at 300 GV. 
Future precision measurements of light and heavy nuclei will provide a critical test of this scenario.

For comparison, we also adopt the CR composition inferred fromt AMS-02 observations at reference rigidities of 100 GV and 1 TV~\cite{AMS:2018qxq, AMS:2018cen, AMS:2015tnn, AMS:2021lxc, AMS:2020cai}. We also consider a scenario where injected CRs follow solar abundances, $f_{\odot}$, as expected if they are swept from the interstellar medium (ISM)~\cite{Lodders:2010kx}. The adopted composition ratios are summarized in Table~\ref{tab:composition}.
As demonstrated in Ref.~\cite{Caprioli:2017oun}, the nonthermal tail of nuclei is enhanced by a factor of $(A/Z)^2$, $f_{\odot}^{\rm enhance}$, for efficient diffusive acceleration in nonrelativistic shocks considering fully ionized state. Our analysis shows that reaccelerating low-energy CRs (100 GV -- 1 TV) yields a mean logarithmic mass $\langle \ln A \rangle$ roughly consistent with observations near the CR knee. In contrast, models assuming solar or enhanced solar compositions fail to reproduce the observed mass distribution. 
For CRs from microquasars, the $(A/Z)^2 \sim A^2$ enhancement scenario --   expected in a singly ionized medium -- predicts a heavier CR composition than a fully ionized medium. However, this leads to an overly heavy composition, inconsistent with the observed $\langle \ln A \rangle$.
These results might indicate that the reacceleration of preexisting low-energy CRs provides a plausible explanation for the composition trends observed in the knee region.

\begin{table}[]
    \caption{Injected CR composition at the rigidity 300 GV, 100 GV, and 1 TV for different composition models. In addition, we also include solar composition $f_{\odot}$ and enhanced solar composition $f_{\odot}^{\rm enhance}$ by a factor of $(A/Z)^2$ for a fully ionized medium.}
    \centering
    \begin{tabular}{lccccccc}
    & Z & A & $f_{A, \rm 300GV}$ & $f_{A, \rm 100GV}$ & $f_{A, \rm 1 TV}$ & $f_{\odot}$  & $f_{\odot}^{\rm enhance}$ \\
       \hline
     P & 1 & 1  & 0.787  & 0.803 & 0.767 & 0.922 & 0.747 \\
     He & 2 & 4 & 0.194 & 0.18 & 0.212 & 0.077  & 0.250 \\
     CNO & 7 & 16 & 0.0145 & 0.013 & 0.0157 & 0.007  & 0.003 \\
     NeMgAlSi & 12 & 28 & 0.00374 & 0.00352 & 0.004 & 0.001  & 0.0006 \\
     Fe & 26 & 56 & 0.00121 & 0.00097 & 0.0014 & 0.00002  & 0.0001
    \end{tabular}
    \label{tab:composition}
\end{table}

\section{Connection to low- and high-energy components}\label{sec:connection}
\begin{figure*}
    \centering
\includegraphics[width=\linewidth]{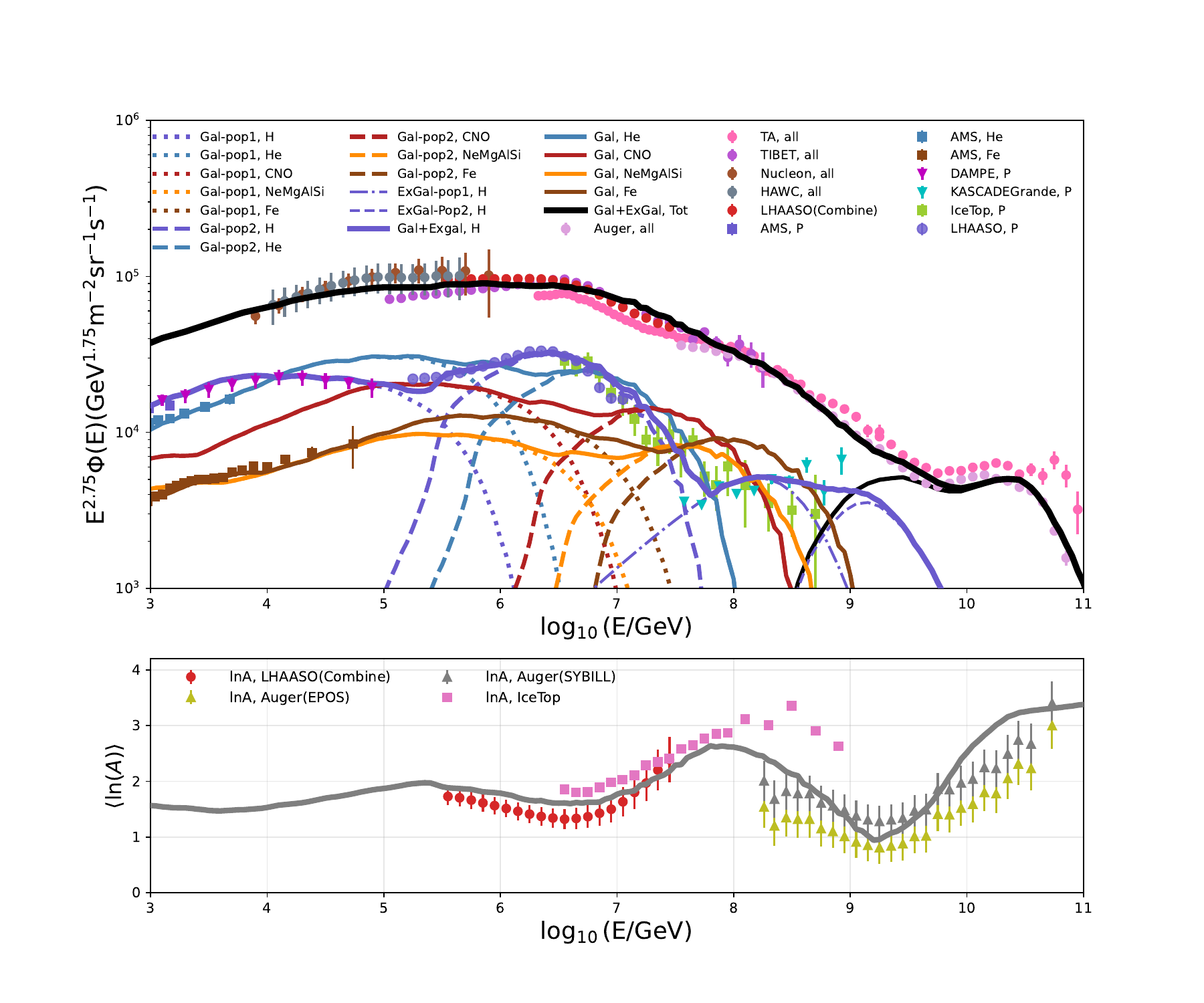}
    \caption{An overall picture of the CR spectrum from Galactic to extragalactic components. \textbf{Gal pop1}: CRs from SNRs~\cite{2008ARA&A..46...89R, Blasi:2013rva} (dotted curve); \textbf{Gal pop2}: CRs from MQs (dashed curve); \textbf{ExGal pop1}: CRs from extragalactic sources, e.g., CGs~\cite{Murase:2008yt} (dotted-dashed curve); \textbf{ExGal pop2}: UHECRs from RGs~\cite{Kimura:2017ubz} (long dashed curve). 
    The data points are the same as those in Fig.~\ref{fig:knee-mq}, with the AMS data taken from Refs.~\cite{AMS:2018qxq, AMS:2018cen, AMS:2015tnn, AMS:2021lxc, AMS:2020cai}, and the mean mass distribution measured by Auger is taken from Ref.~\cite{Yushkov:2020nhr}.
    }
    \label{fig:CR_spectrum_all_lnA}
\end{figure*}

In this section, we aim to demonstrate that our microquasar model can be combined with other astrophysical models explaining the CR spectrum at different wavelengths. In particular, we provide a possible model, in which the whole CR spectrum from around GeV to 100 EeV can be explained with four main sources, including SNRs (Gal pop1), microquasars (MQs, Gal pop2), clusters of galaxies (CGs, ExGal pop1) and radio galaxies (RQs, ExGal pop2), as shown in Fig~\ref{fig:CR_spectrum_all_lnA}.
Note that the main purpose here is to demonstrate that the reacceleration model works not only in the microquasars but also in the large-scale jet-cocoon system in radio galaxies.

For CRs from Galactic SNRs, we calculate the observed energy spectrum of each element with \textsc{DRAGON} code, and the normalization is adjusted to match the AMS-02 data. 
We adopt a broken-power-law injection spectrum of CRs from SNRs to fit the data. See the Appendix for the details.
The escaped CR spectrum typically has an exponential cutoff beyond the maximum acceleration energy.
We set the maximum rigidity from Galactic SNRs to $\mathcal{R}_{\rm max} = 0.4\rm~PV$. 

For CRs in the knee energy range, we use the CR spectrum that comes directly from Monte Carlo simulations, as in Fig.~\ref{fig:knee-mq}. At energies above around 100 PeV to EeV, CRs from extragalactic sources begin to dominate.
At this energy range, the second knee appears, where the spectrum is further steepened from $\gamma \sim 3$ to $\gamma \sim 3.3$, because of the limitation of the maximum energy of the second Galactic component based on the rigidity-dependent modeling.
The physical origin of the CRs beyond the second knee up to the ankle has been proposed by various models, and it has been shown that an additional component so called the B-component is required~\cite{2005JPhG...31R..95H, Gaisser:2006xs}. This component may be responsible for the proton subankle (hardening) feature around 100~PeV observed by KASCADE-Grande~\cite{Apel:2013uni}
, and the proposed models include the cluster of galaxies~\cite{Murase:2008yt,Murase:2013rfa,Simeon:2025gxd}, Galactic wind~\cite{Thoudam:2016syr,Mukhopadhyay:2023kel}, and Wolf-Rayet star explosions~\cite{Thoudam:2016syr}. In addition, superknee CR accelerators such as Galactic neutron star merger remnants~\cite{Kimura:2018ggg} and interacting supernovae~\cite{Ptuskin:2003zv, 2005A&A...429..755P} may explain another hardening feature around $10^{16.22}$~eV reported by TALE~\cite{TelescopeArray:2018bya}. 

For the subankle proton component, we consider a cluster of galaxies for the demonstrative purposes. We consider that the injected CRs have a composition similar to the solar composition, but enhanced by a factor of $(A/Z)^2$; see the last column in Tabel~\ref{tab:composition} about $f_{\odot}^{\rm enhance}$. Note that our results are unaffected by the composition assumption unless the CRs are dominated by heavy nuclei, and such nuclei can be largely disintegrated inside the clusters~\cite{Kotera:2009ms,Fang:2017zjf}. 
The source spectral index above 10-100~PeV is $s_{\rm cr} = 2.4$ and the maximum acceleration rigidity $\mathcal{R}_{\rm max} = 3.7 \times 10^{17}\rm~eV$, following Ref.~\cite{Murase:2008yt}. 
We propagate CRs from an ensemble of extragalactic sources that are uniformly distributed with the characteristic distance between sources $d_s = 32.5\rm~Mpc$. 
The number density of galaxy clusters is around $10^{-5}\rm~Mpc^{-3}$, which leads to $d_s\sim 46$~Mpc.
The observed flux is obtained with Eq.~6 in Ref.~\cite{Gonzalez:2021ajv}.
We assumed a uniform extragalactic magnetic field $B_{\rm rms} = 1\rm~nG$ and coherence length $L_{\rm coh} = 1\rm~Mpc$.

UHECRs above EeV energies are produced from radio galaxies via the shear acceleration mechanism~\cite{Caprioli:2015zka,Kimura:2017ubz, Mbarek:2019glq, Mbarek:2022nat, Seo:2023dkc, Wang:2024ijr, Mbarek:2024nvv}. 
Here, we adopt the predicted spectrum and composition model of UHECRs in KMZ18, where CRs with rigidity $\sim 1$ TV in radio galaxies are reaccelerated by shear in kpc-scale jets. 
Since radio galaxies are more evolved than our Galaxy, KMZ18 considers that the metal composition is enhanced by a ratio of 3, compared to our Galaxy. Here, we further enhance the iron composition ratio by a factor of 3 to improve the fit to the latest UHECR data.
The further enhancement of the iron composition ratio in radio galaxies is reasonable because elliptical galaxies have lower star-formation rates than our Galaxy. In radio galaxies, core-collapse supernovae do not frequently occur because of their low star-formation rates, while type Ia supernovae occur because explosions of white dwarfs require significant time lag after star-formation activities. Then, type Ia supernovae inject iron into the ISM, enhancing the iron composition ratio. 
In addition, we consider a larger value of magnetic field strength in the cocoon with $B_{\rm coc} = 4.5\rm~\mu G$, where the peak energy of the injection spectrum shown in Fig.~3 in KMZ18 is increased to $\mathcal{R}_{\rm max} \approx 2.4\rm~EV$ .

We found that the overall CR spectrum and composition could be explained by the above four populations. This demonstrates that the reacceleration of Galactic CRs plays a key role in reshaping the mass distribution of accelerated CRs at different energy ranges. However, one should keep in mind that uncertainty is currently large.

\section{Summary and Discussion}\label{sec:sum}
In this work, we studied the reacceleration and escape of CRs in microquasars within the framework of the shear acceleration mechanism.
Our results showed that the CRs from microquasars could explain the observed proton energy spectrum above the PeV energy range. 
In our model, the relative fraction of nuclei is determined by the direct measurements of CRs. The predicted all-particle spectrum and composition ratio are consistent with the LHAASO observations if the CR injection rigidity is 100 GV to 1 TV.
The maximum acceleration energy of CRs from shear acceleration in microquasars is close to the knee around 3~PeV, and the escaping CRs have a hard spectrum.

The required number of microquasars in our Galaxy is three to ten (compare Eqs. [\ref{eq:Lcr_shear}] and [\ref{eq:Lcr_tot}]). The lifetime of microquasars in our Galaxy is uncertain, but it might be $0.03-1$ Myr \cite{Yamamoto:2008xr,Sakemi:2023aa}, which is comparable to or longer than the escape time of the Knee CRs ({$\sim$80} kyr).
If the lifetime is comparable or shorter than the CR escape time, the CR density in the ISM might be spatially inhomogeneous and temporary intermittent~\cite{Giacinti:2023upw, Stall:2024klf}.
In addition, we assume that microquasars in our Galaxy have identical CR spectral shape and CR luminosity for simplicity. In reality, there are distributions of CR luminosity and spectral shape, depending on the jet power and jet size of individual microquasars. Investigating the effect of these remains as a future work.

In our scenario, the magnetic field strength in jets and cocoons should be as high as $B_{\rm jet}\sim50 - 100 \rm ~\mu G$ and $B_{\rm coc}\sim10-30\rm~\mu G$ to achieve the PeV CR acceleration. Both field strengths are somewhat higher than expected, but the energetics arguments are not violated. The magnetic luminosity in the jet can be estimated to be 
\begin{align}
 L_B&=\frac{B_{\rm jet}^2}{4\pi} r_{\rm jet}^2 \beta_{\rm jet}c \\
 &\simeq  3.2\times10^{38}\left(\frac{B_{\rm jet}}{\rm 50\mu G}\right)^2\left(\frac{r_{\rm jet}}{\rm 3 ~pc}\right)\left(\frac{\beta_{\rm jet}}{0.3}\right)\rm~erg~s^{-1}.\nonumber
\end{align}
This is slightly lower than the Eddington luminosity, $L_{\rm Edd}\sim10^{39}(M/10~M_\odot)\rm~erg~s^{-1}$. The required cocoon magnetic field strength, $B_{\rm coc}\sim10\rm~\mu G$, is a few times higher than the typical ISM value, $B_{\rm ISM}\sim3\rm~\mu G$, and thus, the shock compression of the ISM would be able to naturally achieve the required $B_{\rm coc}$.

Given the universality of the shear acceleration mechanism, we consider a unified model in which microquasars dominate observed Galactic CRs around knee, while radio galaxies are responsible for the observed UHECRs.
We found that the whole CR range from GeV to 100 EeV could be explained given that SNRs supply most low-energy Galactic CRs and another population (e.g., galaxy clusters) could contribute at subankle energies. 
Our model could be tested with future CR experiments with improved measurements of the composition and anisotropy.

\medskip
We gratefully acknowledge the valuable discussions and insightful comments provided by Walter Winter, Martin Pohl, and Andrew Taylor.
B.T.Z. is supported in China by National Key R\&D program of China under the grant 2024YFA1611402.
This work is partly supported by KAKENHI Nos. 22K14028, 21H04487, 23H04899 (S.S.K.), and 20H05852 (K.M.). S.S.K. acknowledges the support by the Tohoku Initiative for Fostering Global Researchers for Interdisciplinary Sciences (TI-FRIS) of MEXT's Strategic Professional Development Program for Young Researchers. The Monte Carlo simulations of shear acceleration performed by S.S.K. were carried out on Cray XD2000 at the Center for Computational Astrophysics, National Astronomical Observatory of Japan.
K.M. is supported by the NSF Grant Nos. AST-2108466, AST-2108467 and AST-2308021. 
K.M. also acknowledges the SUGAR 2024 workshop in Madison, Wisconsin, which led to the idea of shear acceleration. 

\medskip

\clearpage

\appendix

\section{Dependence on source parameters}\label{app:param}
\begin{figure}
    \centering
\includegraphics[width=1.0\linewidth]{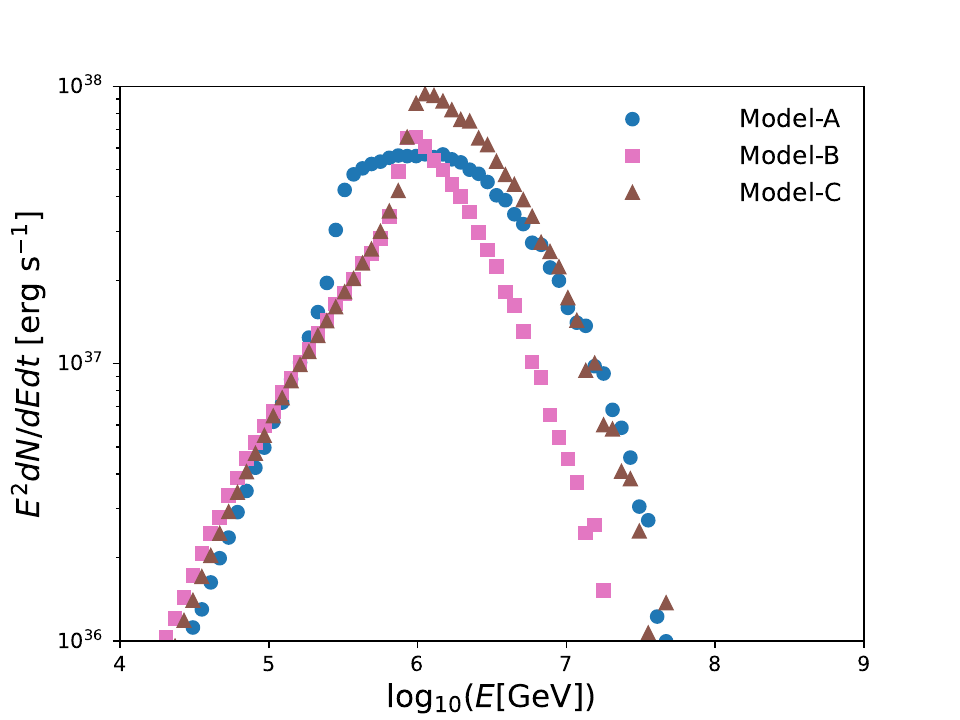}
    \caption{The energy spectrum of CRs predicted in the shear acceleration mechanism from Monte Carlo simulations. Three different parameter sets are considered.  
    }
    \label{fig:spectrum-inj-appendix}
\end{figure}

\begin{figure}
    \centering
    \includegraphics[width=1.\linewidth]{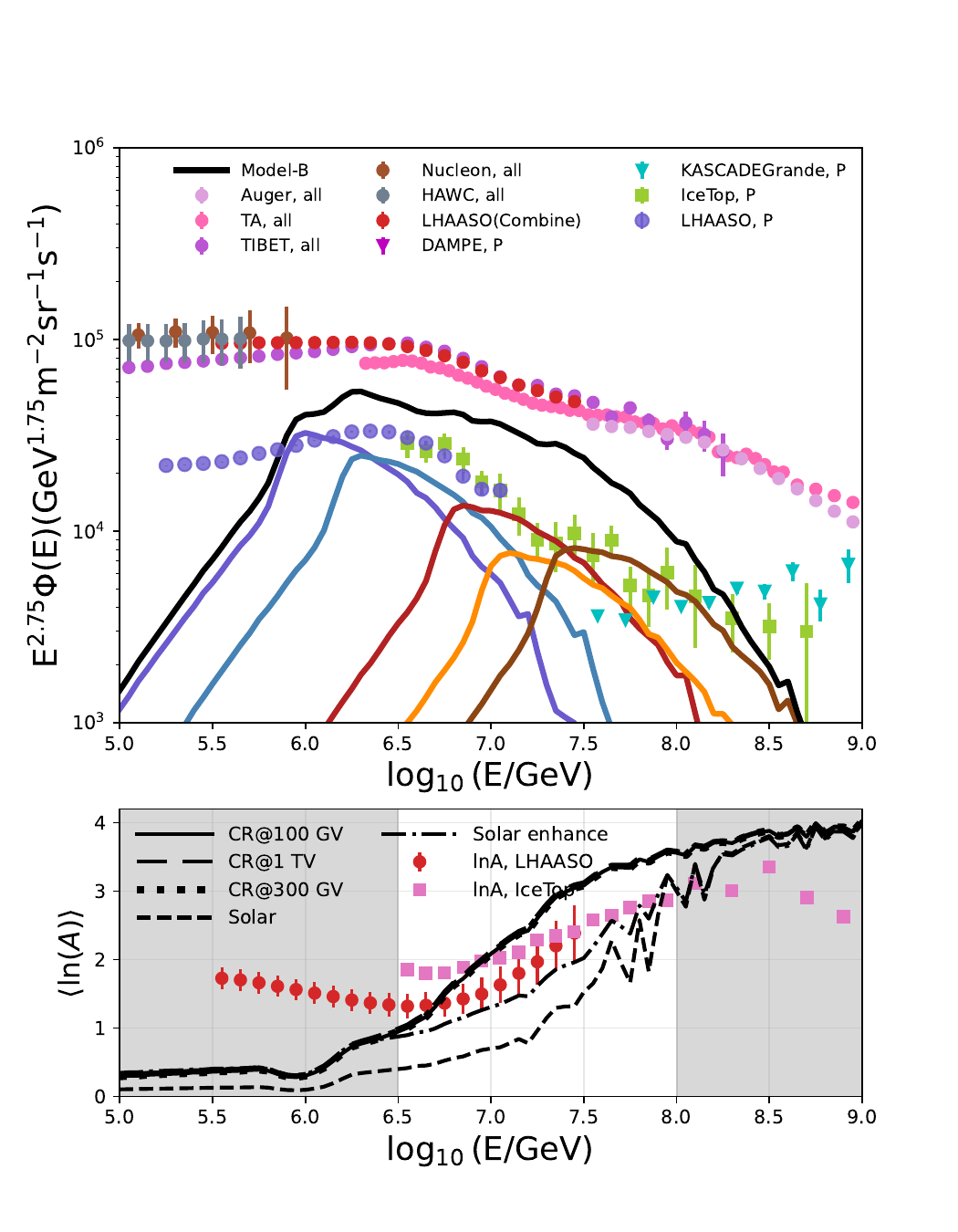}
    \caption{Similar to Fig.~\ref{fig:knee-mq}, but with parameters from Model B.}
    \label{fig:Model-B}
\end{figure}

\begin{figure}
    \centering
    \includegraphics[width=1.\linewidth]{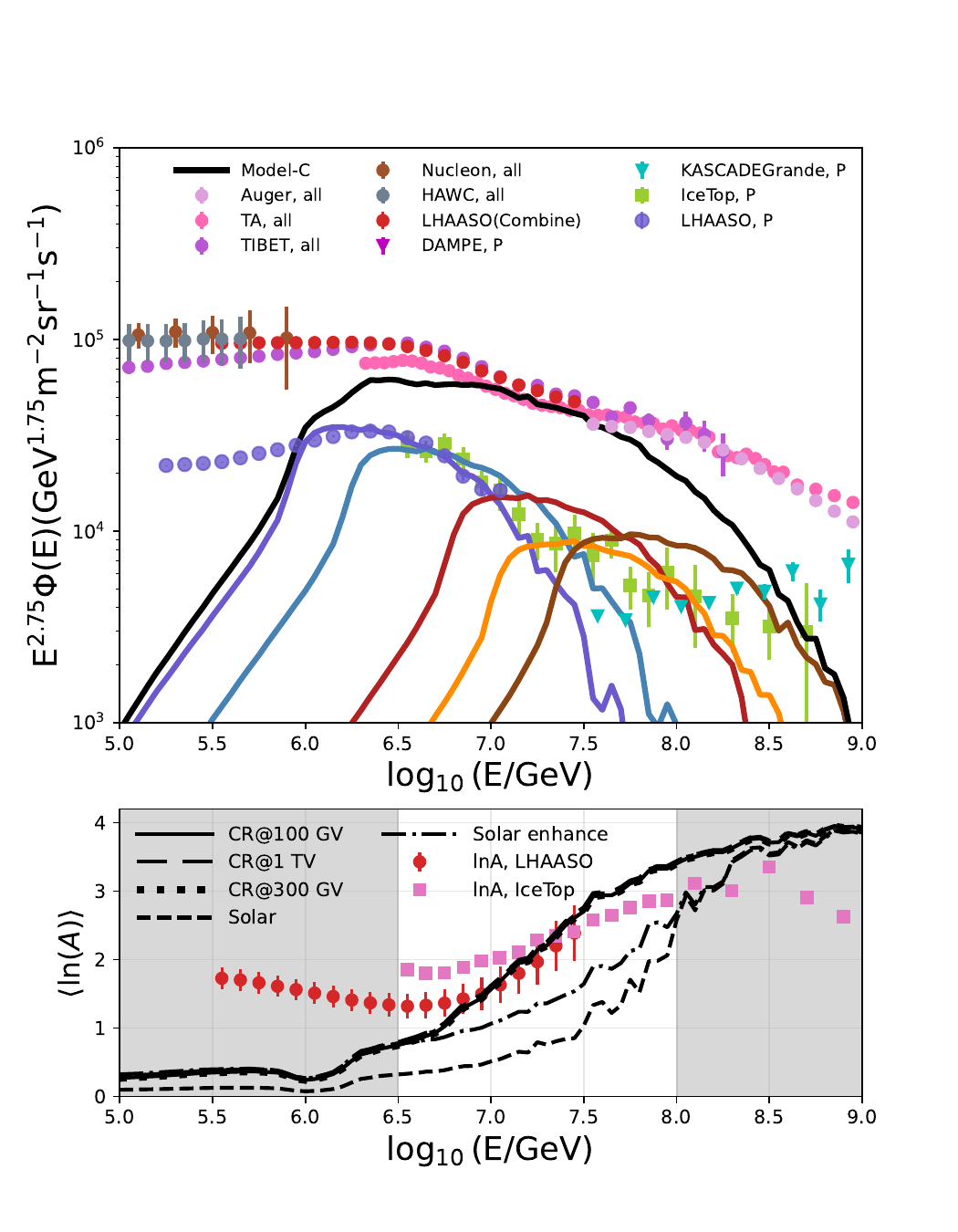}
    \caption{Similar to Fig.~\ref{fig:knee-mq}, but with parameters from Model C.}
    \label{fig:Model-C}
\end{figure}

Here, we present the results for parameters that are different from the default one in Table~\ref{tab:parameters-default}, as indicated in Table~\ref{tab:parameters-all}. 
The predicted energy spectrum of escaping CRs is summarized in Fig.~\ref{fig:spectrum-inj-appendix}.
The predicted CR intensity on Earth is presented in Figs.~\ref{fig:Model-B} and Fig.~\ref{fig:Model-C}, respectively.
We can see the peak energy predicted with Model B is slightly below PeV, which is unable to explain the observed proton knee structure.

\begin{table}[]
    \caption{Model parameters used in Monte Carlo simulations. }
    \centering
    \begin{tabular}{l|ccccccc}
       \hline
    Model & $r_{\rm jet}$ & $B_{\rm jet}$ & $\beta_{\rm jet}$ & $r_{\rm coc}$ & $B_{\rm coc}$ & $l_{\rm coh}$ & $l_{\rm jet}$  \\
     & [pc] & [$\rm \mu G$] &  & [pc] & [$\rm \mu G$] & [pc] & [pc] \\
       \hline
     A (default) & 4 & 50 & 0.2 & 60 & 20  & 1.8 & 100 \\
     B & 3 & 50 & 0.3 & 60 & 10  & 1.8 & 100 \\
     C & 3 & 100 & 0.4 & 60 & 30  & 12 & 100 \\
       \hline
    \end{tabular}
    \label{tab:parameters-all}
\end{table}

\section{Propagate CRs in the Galaxy}\label{app:propa}
\begin{figure}
    \centering
    \includegraphics[width=\linewidth]{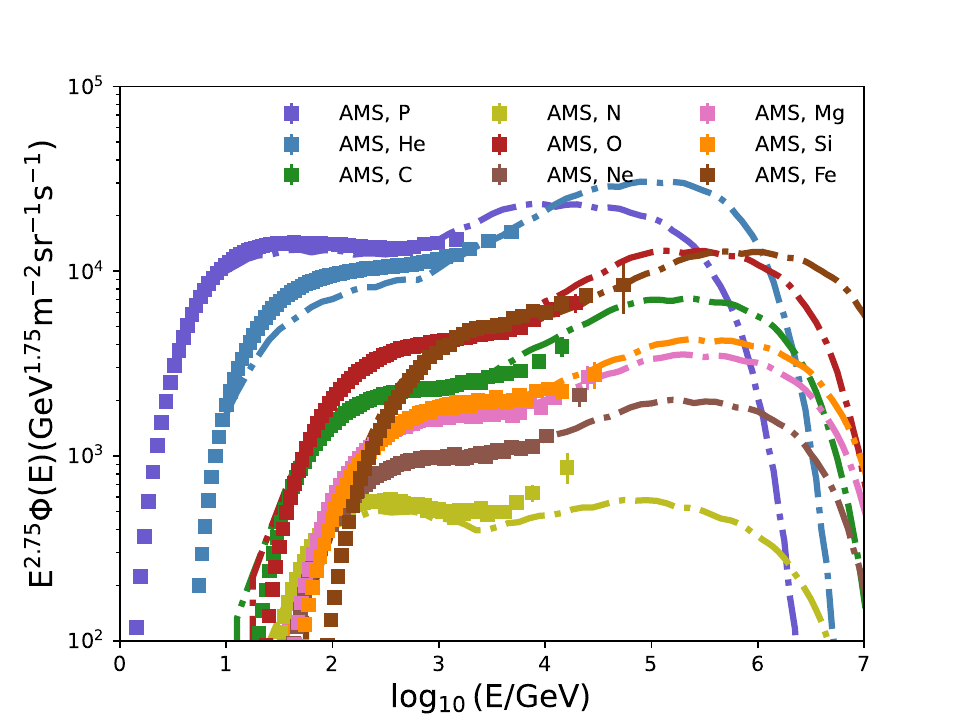}
    \caption{Observed energy spectrum of single element CRs from Galactic SNRs.}
    \label{fig:low-CR}
\end{figure}

In Fig.~\ref{fig:low-CR}, we show the observed energy spectrum of CR species calculated from \textsc{DRAGON} code. 
The injection spectrum from Galactic SNRs is assumed to have a broken powerlaw formula,
\begin{equation}
\frac{dN}{d\mathcal{R}} \propto
\begin{cases}
  \mathcal{R}^{-\gamma_1}{\rm exp}(-\mathcal{R}/\mathcal{R}_{\rm max}), & \mathcal{R} \leq \mathcal{R}_{12} \\
  \mathcal{R}^{-\gamma_2}{\rm exp}(-\mathcal{R}/\mathcal{R}_{\rm max}), & \mathcal{R}_{12} < \mathcal{R} \leq \mathcal{R}_{23} \\
  \mathcal{R}^{-\gamma_3}{\rm exp}(-\mathcal{R}/\mathcal{R}_{\rm max}), & \mathcal{R}_{23} < \mathcal{R} \leq \mathcal{R}_{34} \\
  \mathcal{R}^{-\gamma_4}{\rm exp}(-\mathcal{R}/\mathcal{R}_{\rm max}), & \mathcal{R}_{34} < \mathcal{R} \leq \mathcal{R}_{45} \\
  \mathcal{R}^{-\gamma_5} {\rm exp}(-\mathcal{R}/\mathcal{R}_{\rm max}), & \mathcal{R} > \mathcal{R}_{45}
\end{cases}
\end{equation}
where $\mathcal{R}_{\rm max} = 0.4\rm~PV$. The adopted parameters are listed in Table~\ref{tab:DRAGON_param} and the energy spectrum is presented in Fig.~\ref{fig:low-CR}.
\begin{table}[]
    \centering
    \caption{Parameters used in \textsc{DRAGON} for the propagation of CRs from Galactic SNRs. Note that the parameters for the nuclei that were heavier than neon were keep the same, and not listed here.}
    \begin{tabular}{lcccccc}
    \hline
    & H & He & C & N & O & Ne \\
    \hline
        $\gamma_1$ &  1.8 & 2 & 2 & 1.9 & 2 & 1.9\\
         $\gamma_2$ & 2.4 & 2.28 & 2.35 & 2.4 & 2.38 & 2.4 \\
          $\gamma_3$ & 2.26 & 2.16 & 2.15 & 2.24 & 2.15 & 2.24\\
           $\gamma_4$ & 2.16 & 2.1 & 2.15 & 2.24 & 2.15 & 2.24\\
            $\gamma_5$ & 2.4 & 2.28 & 2.35 & 2.4 & 2.38 & 2.4\\
            $\mathcal{R}_{12}$ & 7 & 7 & 7 & 7 & 7 & 7\\
            $\mathcal{R}_{23}$ & 335 & 165 &  165 & 330 & 165 & 330\\
            $\mathcal{R}_{34}$ & 1000 & 1000 & 1000 & 1000 & 1000 & 1000\\
            $\mathcal{R}_{45}$ & 5000 & 5000 & 5000 & 5000 & 5000 & 5000\\
         & 
    \end{tabular}
    \label{tab:DRAGON_param}
\end{table}

\bibliography{bzhang}

\providecommand{\noopsort}[1]{}\providecommand{\singleletter}[1]{#1}%
\begin{thebibliography}{119}%
\makeatletter
\providecommand \@ifxundefined [1]{%
 \@ifx{#1\undefined}
}%
\providecommand \@ifnum [1]{%
 \ifnum #1\expandafter \@firstoftwo
 \else \expandafter \@secondoftwo
 \fi
}%
\providecommand \@ifx [1]{%
 \ifx #1\expandafter \@firstoftwo
 \else \expandafter \@secondoftwo
 \fi
}%
\providecommand \natexlab [1]{#1}%
\providecommand \enquote  [1]{``#1''}%
\providecommand \bibnamefont  [1]{#1}%
\providecommand \bibfnamefont [1]{#1}%
\providecommand \citenamefont [1]{#1}%
\providecommand \href@noop [0]{\@secondoftwo}%
\providecommand \href [0]{\begingroup \@sanitize@url \@href}%
\providecommand \@href[1]{\@@startlink{#1}\@@href}%
\providecommand \@@href[1]{\endgroup#1\@@endlink}%
\providecommand \@sanitize@url [0]{\catcode `\\12\catcode `\$12\catcode
  `\&12\catcode `\#12\catcode `\^12\catcode `\_12\catcode `\%12\relax}%
\providecommand \@@startlink[1]{}%
\providecommand \@@endlink[0]{}%
\providecommand \url  [0]{\begingroup\@sanitize@url \@url }%
\providecommand \@url [1]{\endgroup\@href {#1}{\urlprefix }}%
\providecommand \urlprefix  [0]{URL }%
\providecommand \Eprint [0]{\href }%
\providecommand \doibase [0]{http://dx.doi.org/}%
\providecommand \selectlanguage [0]{\@gobble}%
\providecommand \bibinfo  [0]{\@secondoftwo}%
\providecommand \bibfield  [0]{\@secondoftwo}%
\providecommand \translation [1]{[#1]}%
\providecommand \BibitemOpen [0]{}%
\providecommand \bibitemStop [0]{}%
\providecommand \bibitemNoStop [0]{.\EOS\space}%
\providecommand \EOS [0]{\spacefactor3000\relax}%
\providecommand \BibitemShut  [1]{\csname bibitem#1\endcsname}%
\let\auto@bib@innerbib\@empty
\bibitem [{\citenamefont {Antoni}\ \emph {et~al.}(2005)\citenamefont {Antoni}
  \emph {et~al.}}]{KASCADE:2005ynk}%
  \BibitemOpen
  \bibfield  {author} {\bibinfo {author} {\bibfnamefont {T.}~\bibnamefont
  {Antoni}} \emph {et~al.} (\bibinfo {collaboration} {KASCADE}),\ }\href
  {\doibase 10.1016/j.astropartphys.2005.04.001} {\bibfield  {journal}
  {\bibinfo  {journal} {Astropart. Phys.}\ }\textbf {\bibinfo {volume} {24}},\
  \bibinfo {pages} {1} (\bibinfo {year} {2005})},\ \Eprint
  {http://arxiv.org/abs/astro-ph/0505413} {arXiv:astro-ph/0505413} \BibitemShut
  {NoStop}%
\bibitem [{\citenamefont {Kotera}\ and\ \citenamefont
  {Olinto}(2011)}]{Kotera:2011cp}%
  \BibitemOpen
  \bibfield  {author} {\bibinfo {author} {\bibfnamefont {K.}~\bibnamefont
  {Kotera}}\ and\ \bibinfo {author} {\bibfnamefont {A.~V.}\ \bibnamefont
  {Olinto}},\ }\href {\doibase 10.1146/annurev-astro-081710-102620} {\bibfield
  {journal} {\bibinfo  {journal} {Ann. Rev. Astron. Astrophys.}\ }\textbf
  {\bibinfo {volume} {49}},\ \bibinfo {pages} {119} (\bibinfo {year} {2011})},\
  \Eprint {http://arxiv.org/abs/1101.4256} {arXiv:1101.4256 [astro-ph.HE]}
  \BibitemShut {NoStop}%
\bibitem [{\citenamefont {Kachelriess}\ and\ \citenamefont
  {Semikoz}(2019)}]{Kachelriess:2019oqu}%
  \BibitemOpen
  \bibfield  {author} {\bibinfo {author} {\bibfnamefont {M.}~\bibnamefont
  {Kachelriess}}\ and\ \bibinfo {author} {\bibfnamefont {D.~V.}\ \bibnamefont
  {Semikoz}},\ }\href {\doibase 10.1016/j.ppnp.2019.07.002} {\bibfield
  {journal} {\bibinfo  {journal} {Prog. Part. Nucl. Phys.}\ }\textbf {\bibinfo
  {volume} {109}},\ \bibinfo {pages} {103710} (\bibinfo {year} {2019})},\
  \Eprint {http://arxiv.org/abs/1904.08160} {arXiv:1904.08160 [astro-ph.HE]}
  \BibitemShut {NoStop}%
\bibitem [{\citenamefont {Amenomori}\ \emph {et~al.}(2008)\citenamefont
  {Amenomori} \emph {et~al.}}]{TIBETIII:2008qon}%
  \BibitemOpen
  \bibfield  {author} {\bibinfo {author} {\bibfnamefont {M.}~\bibnamefont
  {Amenomori}} \emph {et~al.} (\bibinfo {collaboration} {TIBET III}),\ }\href
  {\doibase 10.1086/529514} {\bibfield  {journal} {\bibinfo  {journal}
  {Astrophys. J.}\ }\textbf {\bibinfo {volume} {678}},\ \bibinfo {pages} {1165}
  (\bibinfo {year} {2008})},\ \Eprint {http://arxiv.org/abs/0801.1803}
  {arXiv:0801.1803 [hep-ex]} \BibitemShut {NoStop}%
\bibitem [{\citenamefont {Cao}\ \emph {et~al.}(2024{\natexlab{a}})\citenamefont
  {Cao} \emph {et~al.}}]{LHAASO:2024knt}%
  \BibitemOpen
  \bibfield  {author} {\bibinfo {author} {\bibfnamefont {Z.}~\bibnamefont
  {Cao}} \emph {et~al.} (\bibinfo {collaboration} {LHAASO}),\ }\href {\doibase
  10.1103/PhysRevLett.132.131002} {\bibfield  {journal} {\bibinfo  {journal}
  {Phys. Rev. Lett.}\ }\textbf {\bibinfo {volume} {132}},\ \bibinfo {pages}
  {131002} (\bibinfo {year} {2024}{\natexlab{a}})},\ \Eprint
  {http://arxiv.org/abs/2403.10010} {arXiv:2403.10010 [astro-ph.HE]}
  \BibitemShut {NoStop}%
\bibitem [{\citenamefont {Kulikov}\ and\ \citenamefont
  {Khristiansen}(1958)}]{Kulikov1958}%
  \BibitemOpen
  \bibfield  {author} {\bibinfo {author} {\bibfnamefont {G.~V.}\ \bibnamefont
  {Kulikov}}\ and\ \bibinfo {author} {\bibfnamefont {G.~B.}\ \bibnamefont
  {Khristiansen}},\ }\href
  {http://jetp.ras.ru/cgibin/e/index/e/8/3/p441?a=list} {\bibfield  {journal}
  {\bibinfo  {journal} {Journal of Experimental and Theoretical Physics}\
  }\textbf {\bibinfo {volume} {35}},\ \bibinfo {pages} {8} (\bibinfo {year}
  {1958})}\BibitemShut {NoStop}%
\bibitem [{\citenamefont {Nagano}\ \emph {et~al.}(1992)\citenamefont {Nagano},
  \citenamefont {Teshima}, \citenamefont {Matsubara}, \citenamefont {Dai},
  \citenamefont {Hara}, \citenamefont {Hayashida}, \citenamefont {Honda},
  \citenamefont {Ohoka},\ and\ \citenamefont {Yoshida}}]{Nagano:1991jz}%
  \BibitemOpen
  \bibfield  {author} {\bibinfo {author} {\bibfnamefont {M.}~\bibnamefont
  {Nagano}}, \bibinfo {author} {\bibfnamefont {M.}~\bibnamefont {Teshima}},
  \bibinfo {author} {\bibfnamefont {Y.}~\bibnamefont {Matsubara}}, \bibinfo
  {author} {\bibfnamefont {H.~Y.}\ \bibnamefont {Dai}}, \bibinfo {author}
  {\bibfnamefont {T.}~\bibnamefont {Hara}}, \bibinfo {author} {\bibfnamefont
  {N.}~\bibnamefont {Hayashida}}, \bibinfo {author} {\bibfnamefont
  {M.}~\bibnamefont {Honda}}, \bibinfo {author} {\bibfnamefont
  {H.}~\bibnamefont {Ohoka}}, \ and\ \bibinfo {author} {\bibfnamefont
  {S.}~\bibnamefont {Yoshida}},\ }\href {\doibase 10.1088/0954-3899/18/2/022}
  {\bibfield  {journal} {\bibinfo  {journal} {J. Phys. G}\ }\textbf {\bibinfo
  {volume} {18}},\ \bibinfo {pages} {423} (\bibinfo {year} {1992})}\BibitemShut
  {NoStop}%
\bibitem [{\citenamefont {Abbasi}\ \emph {et~al.}(2018)\citenamefont {Abbasi}
  \emph {et~al.}}]{TelescopeArray:2018bya}%
  \BibitemOpen
  \bibfield  {author} {\bibinfo {author} {\bibfnamefont {R.~U.}\ \bibnamefont
  {Abbasi}} \emph {et~al.} (\bibinfo {collaboration} {Telescope Array}),\
  }\href {\doibase 10.3847/1538-4357/aada05} {\bibfield  {journal} {\bibinfo
  {journal} {Astrophys. J.}\ }\textbf {\bibinfo {volume} {865}},\ \bibinfo
  {pages} {74} (\bibinfo {year} {2018})},\ \Eprint
  {http://arxiv.org/abs/1803.01288} {arXiv:1803.01288 [astro-ph.HE]}
  \BibitemShut {NoStop}%
\bibitem [{\citenamefont {Bird}\ \emph {et~al.}(1993)\citenamefont {Bird} \emph
  {et~al.}}]{HiRes:1993hwv}%
  \BibitemOpen
  \bibfield  {author} {\bibinfo {author} {\bibfnamefont {D.~J.}\ \bibnamefont
  {Bird}} \emph {et~al.} (\bibinfo {collaboration} {HiRes}),\ }\href {\doibase
  10.1103/PhysRevLett.71.3401} {\bibfield  {journal} {\bibinfo  {journal}
  {Phys. Rev. Lett.}\ }\textbf {\bibinfo {volume} {71}},\ \bibinfo {pages}
  {3401} (\bibinfo {year} {1993})}\BibitemShut {NoStop}%
\bibitem [{\citenamefont {Abbasi}\ \emph {et~al.}(2008)\citenamefont {Abbasi}
  \emph {et~al.}}]{Abbasi:2007sv}%
  \BibitemOpen
  \bibfield  {author} {\bibinfo {author} {\bibfnamefont {R.~U.}\ \bibnamefont
  {Abbasi}} \emph {et~al.} (\bibinfo {collaboration} {HiRes Collaboration}),\
  }\href {\doibase 10.1103/PhysRevLett.100.101101} {\bibfield  {journal}
  {\bibinfo  {journal} {Phys. Rev. Lett.}\ }\textbf {\bibinfo {volume} {100}},\
  \bibinfo {pages} {101101} (\bibinfo {year} {2008})},\ \Eprint
  {http://arxiv.org/abs/astro-ph/0703099} {arXiv:astro-ph/0703099 [astro-ph]}
  \BibitemShut {NoStop}%
\bibitem [{\citenamefont {Abraham}\ \emph {et~al.}(2008)\citenamefont {Abraham}
  \emph {et~al.}}]{Abraham:2008ru}%
  \BibitemOpen
  \bibfield  {author} {\bibinfo {author} {\bibfnamefont {J.}~\bibnamefont
  {Abraham}} \emph {et~al.} (\bibinfo {collaboration} {Pierre Auger
  Collaboration}),\ }\href {\doibase 10.1103/PhysRevLett.101.061101} {\bibfield
   {journal} {\bibinfo  {journal} {Phys. Rev. Lett.}\ }\textbf {\bibinfo
  {volume} {101}},\ \bibinfo {pages} {061101} (\bibinfo {year} {2008})},\
  \Eprint {http://arxiv.org/abs/0806.4302} {arXiv:0806.4302 [astro-ph]}
  \BibitemShut {NoStop}%
\bibitem [{\citenamefont {{Hillas}}(2005)}]{2005JPhG...31R..95H}%
  \BibitemOpen
  \bibfield  {author} {\bibinfo {author} {\bibfnamefont {A.~M.}\ \bibnamefont
  {{Hillas}}},\ }\href {\doibase 10.1088/0954-3899/31/5/R02} {\bibfield
  {journal} {\bibinfo  {journal} {Journal of Physics G Nuclear Physics}\
  }\textbf {\bibinfo {volume} {31}},\ \bibinfo {pages} {R95} (\bibinfo {year}
  {2005})}\BibitemShut {NoStop}%
\bibitem [{\citenamefont {Cardillo}\ \emph {et~al.}(2015)\citenamefont
  {Cardillo}, \citenamefont {Amato},\ and\ \citenamefont
  {Blasi}}]{Cardillo:2015zda}%
  \BibitemOpen
  \bibfield  {author} {\bibinfo {author} {\bibfnamefont {M.}~\bibnamefont
  {Cardillo}}, \bibinfo {author} {\bibfnamefont {E.}~\bibnamefont {Amato}}, \
  and\ \bibinfo {author} {\bibfnamefont {P.}~\bibnamefont {Blasi}},\ }\href
  {\doibase 10.1016/j.astropartphys.2015.03.002} {\bibfield  {journal}
  {\bibinfo  {journal} {Astropart. Phys.}\ }\textbf {\bibinfo {volume} {69}},\
  \bibinfo {pages} {1} (\bibinfo {year} {2015})},\ \Eprint
  {http://arxiv.org/abs/1503.03001} {arXiv:1503.03001 [astro-ph.HE]}
  \BibitemShut {NoStop}%
\bibitem [{\citenamefont {Giacinti}\ \emph {et~al.}(2015)\citenamefont
  {Giacinti}, \citenamefont {Kachelrie\ss{}},\ and\ \citenamefont
  {Semikoz}}]{Giacinti:2015hva}%
  \BibitemOpen
  \bibfield  {author} {\bibinfo {author} {\bibfnamefont {G.}~\bibnamefont
  {Giacinti}}, \bibinfo {author} {\bibfnamefont {M.}~\bibnamefont
  {Kachelrie\ss{}}}, \ and\ \bibinfo {author} {\bibfnamefont {D.~V.}\
  \bibnamefont {Semikoz}},\ }\href {\doibase 10.1103/PhysRevD.91.083009}
  {\bibfield  {journal} {\bibinfo  {journal} {Phys. Rev. D}\ }\textbf {\bibinfo
  {volume} {91}},\ \bibinfo {pages} {083009} (\bibinfo {year} {2015})},\
  \Eprint {http://arxiv.org/abs/1502.01608} {arXiv:1502.01608 [astro-ph.HE]}
  \BibitemShut {NoStop}%
\bibitem [{\citenamefont {Gaisser}(2006)}]{Gaisser:2006xs}%
  \BibitemOpen
  \bibfield  {author} {\bibinfo {author} {\bibfnamefont {T.~K.}\ \bibnamefont
  {Gaisser}},\ }\href {\doibase 10.1088/1742-6596/47/1/002} {\bibfield
  {journal} {\bibinfo  {journal} {J. Phys. Conf. Ser.}\ }\textbf {\bibinfo
  {volume} {47}},\ \bibinfo {pages} {15} (\bibinfo {year} {2006})}\BibitemShut
  {NoStop}%
\bibitem [{\citenamefont {Aab}\ \emph {et~al.}(2018)\citenamefont {Aab} \emph
  {et~al.}}]{PierreAuger:2018zqu}%
  \BibitemOpen
  \bibfield  {author} {\bibinfo {author} {\bibfnamefont {A.}~\bibnamefont
  {Aab}} \emph {et~al.} (\bibinfo {collaboration} {Pierre Auger}),\ }\href
  {\doibase 10.3847/1538-4357/aae689} {\bibfield  {journal} {\bibinfo
  {journal} {Astrophys. J.}\ }\textbf {\bibinfo {volume} {868}},\ \bibinfo
  {pages} {4} (\bibinfo {year} {2018})},\ \Eprint
  {http://arxiv.org/abs/1808.03579} {arXiv:1808.03579 [astro-ph.HE]}
  \BibitemShut {NoStop}%
\bibitem [{\citenamefont {Drury}(2018)}]{Drury:2018mkl}%
  \BibitemOpen
  \bibfield  {author} {\bibinfo {author} {\bibfnamefont {L.~O.}\ \bibnamefont
  {Drury}},\ }\href {\doibase 10.1016/j.nuclphysbps.2018.07.002} {\bibfield
  {journal} {\bibinfo  {journal} {Nucl. Part. Phys. Proc.}\ }\textbf {\bibinfo
  {volume} {297-299}},\ \bibinfo {pages} {6} (\bibinfo {year}
  {2018})}\BibitemShut {NoStop}%
\bibitem [{\citenamefont {Cao}\ \emph {et~al.}(2025)\citenamefont {Cao} \emph
  {et~al.}}]{LHAASO:2025byy}%
  \BibitemOpen
  \bibfield  {author} {\bibinfo {author} {\bibfnamefont {Z.}~\bibnamefont
  {Cao}} \emph {et~al.} (\bibinfo {collaboration} {LHAASO}),\ }\href@noop {} {\
   (\bibinfo {year} {2025})},\ \Eprint {http://arxiv.org/abs/2505.14447}
  {arXiv:2505.14447 [astro-ph.HE]} \BibitemShut {NoStop}%
\bibitem [{\citenamefont {{Reynolds}}(2008)}]{2008ARA&A..46...89R}%
  \BibitemOpen
  \bibfield  {author} {\bibinfo {author} {\bibfnamefont {S.~P.}\ \bibnamefont
  {{Reynolds}}},\ }\href {\doibase 10.1146/annurev.astro.46.060407.145237}
  {\bibfield  {journal} {\bibinfo  {journal} {Annual Review of Astronomy and
  Astrophysics}\ }\textbf {\bibinfo {volume} {46}},\ \bibinfo {pages} {89}
  (\bibinfo {year} {2008})}\BibitemShut {NoStop}%
\bibitem [{\citenamefont {Blasi}(2013)}]{Blasi:2013rva}%
  \BibitemOpen
  \bibfield  {author} {\bibinfo {author} {\bibfnamefont {P.}~\bibnamefont
  {Blasi}},\ }\href {\doibase 10.1007/s00159-013-0070-7} {\bibfield  {journal}
  {\bibinfo  {journal} {Astron. Astrophys. Rev.}\ }\textbf {\bibinfo {volume}
  {21}},\ \bibinfo {pages} {70} (\bibinfo {year} {2013})},\ \Eprint
  {http://arxiv.org/abs/1311.7346} {arXiv:1311.7346 [astro-ph.HE]} \BibitemShut
  {NoStop}%
\bibitem [{\citenamefont {{Bell}}(2004)}]{2004MNRAS.353..550B}%
  \BibitemOpen
  \bibfield  {author} {\bibinfo {author} {\bibfnamefont {A.~R.}\ \bibnamefont
  {{Bell}}},\ }\href {\doibase 10.1111/j.1365-2966.2004.08097.x} {\bibfield
  {journal} {\bibinfo  {journal} {Mon. Not. Roy. Astron. Soc.}\ }\textbf
  {\bibinfo {volume} {353}},\ \bibinfo {pages} {550} (\bibinfo {year}
  {2004})}\BibitemShut {NoStop}%
\bibitem [{\citenamefont {{Bell}}(2005)}]{2005MNRAS.358..181B}%
  \BibitemOpen
  \bibfield  {author} {\bibinfo {author} {\bibfnamefont {A.~R.}\ \bibnamefont
  {{Bell}}},\ }\href {\doibase 10.1111/j.1365-2966.2005.08774.x} {\bibfield
  {journal} {\bibinfo  {journal} {Mon. Not. Roy. Astron. Soc.}\ }\textbf
  {\bibinfo {volume} {358}},\ \bibinfo {pages} {181} (\bibinfo {year}
  {2005})}\BibitemShut {NoStop}%
\bibitem [{\citenamefont {Gabici}\ \emph {et~al.}(2019)\citenamefont {Gabici},
  \citenamefont {Evoli}, \citenamefont {Gaggero}, \citenamefont {Lipari},
  \citenamefont {Mertsch}, \citenamefont {Orlando}, \citenamefont {Strong},\
  and\ \citenamefont {Vittino}}]{Gabici:2019jvz}%
  \BibitemOpen
  \bibfield  {author} {\bibinfo {author} {\bibfnamefont {S.}~\bibnamefont
  {Gabici}}, \bibinfo {author} {\bibfnamefont {C.}~\bibnamefont {Evoli}},
  \bibinfo {author} {\bibfnamefont {D.}~\bibnamefont {Gaggero}}, \bibinfo
  {author} {\bibfnamefont {P.}~\bibnamefont {Lipari}}, \bibinfo {author}
  {\bibfnamefont {P.}~\bibnamefont {Mertsch}}, \bibinfo {author} {\bibfnamefont
  {E.}~\bibnamefont {Orlando}}, \bibinfo {author} {\bibfnamefont
  {A.}~\bibnamefont {Strong}}, \ and\ \bibinfo {author} {\bibfnamefont
  {A.}~\bibnamefont {Vittino}},\ }\href {\doibase 10.1142/S0218271819300222}
  {\bibfield  {journal} {\bibinfo  {journal} {Int. J. Mod. Phys. D}\ }\textbf
  {\bibinfo {volume} {28}},\ \bibinfo {pages} {1930022} (\bibinfo {year}
  {2019})},\ \Eprint {http://arxiv.org/abs/1903.11584} {arXiv:1903.11584
  [astro-ph.HE]} \BibitemShut {NoStop}%
\bibitem [{\citenamefont {Giuliani}\ and\ \citenamefont
  {Cardillo}(2024)}]{Giuliani:2024fpq}%
  \BibitemOpen
  \bibfield  {author} {\bibinfo {author} {\bibfnamefont {A.}~\bibnamefont
  {Giuliani}}\ and\ \bibinfo {author} {\bibfnamefont {M.}~\bibnamefont
  {Cardillo}},\ }\href {\doibase 10.3390/universe10050203} {\bibfield
  {journal} {\bibinfo  {journal} {Universe}\ }\textbf {\bibinfo {volume}
  {10}},\ \bibinfo {pages} {203} (\bibinfo {year} {2024})},\ \Eprint
  {http://arxiv.org/abs/2405.17384} {arXiv:2405.17384 [astro-ph.HE]}
  \BibitemShut {NoStop}%
\bibitem [{\citenamefont {{Casse}}\ and\ \citenamefont
  {{Paul}}(1982)}]{1982ApJ...258..860C}%
  \BibitemOpen
  \bibfield  {author} {\bibinfo {author} {\bibfnamefont {M.}~\bibnamefont
  {{Casse}}}\ and\ \bibinfo {author} {\bibfnamefont {J.~A.}\ \bibnamefont
  {{Paul}}},\ }\href {\doibase 10.1086/160132} {\bibfield  {journal} {\bibinfo
  {journal} {\apj}\ }\textbf {\bibinfo {volume} {258}},\ \bibinfo {pages} {860}
  (\bibinfo {year} {1982})}\BibitemShut {NoStop}%
\bibitem [{\citenamefont {{Cesarsky}}\ and\ \citenamefont
  {{Montmerle}}(1983)}]{1983SSRv...36..173C}%
  \BibitemOpen
  \bibfield  {author} {\bibinfo {author} {\bibfnamefont {C.~J.}\ \bibnamefont
  {{Cesarsky}}}\ and\ \bibinfo {author} {\bibfnamefont {T.}~\bibnamefont
  {{Montmerle}}},\ }\href {\doibase 10.1007/BF00167503} {\bibfield  {journal}
  {\bibinfo  {journal} {Space Sci. Rev.}\ }\textbf {\bibinfo {volume} {36}},\
  \bibinfo {pages} {173} (\bibinfo {year} {1983})}\BibitemShut {NoStop}%
\bibitem [{\citenamefont {Aharonian}\ \emph {et~al.}(2019)\citenamefont
  {Aharonian}, \citenamefont {Yang},\ and\ \citenamefont {de~O\~na
  Wilhelmi}}]{Aharonian:2018oau}%
  \BibitemOpen
  \bibfield  {author} {\bibinfo {author} {\bibfnamefont {F.}~\bibnamefont
  {Aharonian}}, \bibinfo {author} {\bibfnamefont {R.}~\bibnamefont {Yang}}, \
  and\ \bibinfo {author} {\bibfnamefont {E.}~\bibnamefont {de~O\~na
  Wilhelmi}},\ }\href {\doibase 10.1038/s41550-019-0724-0} {\bibfield
  {journal} {\bibinfo  {journal} {Nature Astron.}\ }\textbf {\bibinfo {volume}
  {3}},\ \bibinfo {pages} {561} (\bibinfo {year} {2019})},\ \Eprint
  {http://arxiv.org/abs/1804.02331} {arXiv:1804.02331 [astro-ph.HE]}
  \BibitemShut {NoStop}%
\bibitem [{\citenamefont {Bednarek}\ and\ \citenamefont
  {Protheroe}(2002)}]{Bednarek:2001av}%
  \BibitemOpen
  \bibfield  {author} {\bibinfo {author} {\bibfnamefont {W.}~\bibnamefont
  {Bednarek}}\ and\ \bibinfo {author} {\bibfnamefont {R.~J.}\ \bibnamefont
  {Protheroe}},\ }\href {\doibase 10.1016/S0927-6505(01)00124-4} {\bibfield
  {journal} {\bibinfo  {journal} {Astropart. Phys.}\ }\textbf {\bibinfo
  {volume} {16}},\ \bibinfo {pages} {397} (\bibinfo {year} {2002})},\ \Eprint
  {http://arxiv.org/abs/astro-ph/0103160} {arXiv:astro-ph/0103160} \BibitemShut
  {NoStop}%
\bibitem [{\citenamefont {{Giller}}\ and\ \citenamefont
  {{Lipski}}(2002)}]{2002JPhG...28.1275G}%
  \BibitemOpen
  \bibfield  {author} {\bibinfo {author} {\bibfnamefont {M.}~\bibnamefont
  {{Giller}}}\ and\ \bibinfo {author} {\bibfnamefont {M.}~\bibnamefont
  {{Lipski}}},\ }\href {\doibase 10.1088/0954-3899/28/6/310} {\bibfield
  {journal} {\bibinfo  {journal} {Journal of Physics G Nuclear Physics}\
  }\textbf {\bibinfo {volume} {28}},\ \bibinfo {pages} {1275} (\bibinfo {year}
  {2002})}\BibitemShut {NoStop}%
\bibitem [{\citenamefont {Bednarek}\ and\ \citenamefont
  {Bartosik}(2004)}]{Bednarek:2004wp}%
  \BibitemOpen
  \bibfield  {author} {\bibinfo {author} {\bibfnamefont {W.}~\bibnamefont
  {Bednarek}}\ and\ \bibinfo {author} {\bibfnamefont {M.}~\bibnamefont
  {Bartosik}},\ }\href {\doibase 10.1051/0004-6361:20047005} {\bibfield
  {journal} {\bibinfo  {journal} {Astron. Astrophys.}\ }\textbf {\bibinfo
  {volume} {423}},\ \bibinfo {pages} {405} (\bibinfo {year} {2004})},\ \Eprint
  {http://arxiv.org/abs/astro-ph/0405310} {arXiv:astro-ph/0405310} \BibitemShut
  {NoStop}%
\bibitem [{\citenamefont {Ohira}\ \emph {et~al.}(2018)\citenamefont {Ohira},
  \citenamefont {Kisaka},\ and\ \citenamefont {Yamazaki}}]{Ohira:2017bxa}%
  \BibitemOpen
  \bibfield  {author} {\bibinfo {author} {\bibfnamefont {Y.}~\bibnamefont
  {Ohira}}, \bibinfo {author} {\bibfnamefont {S.}~\bibnamefont {Kisaka}}, \
  and\ \bibinfo {author} {\bibfnamefont {R.}~\bibnamefont {Yamazaki}},\ }\href
  {\doibase 10.1093/mnras/sty1159} {\bibfield  {journal} {\bibinfo  {journal}
  {Mon. Not. Roy. Astron. Soc.}\ }\textbf {\bibinfo {volume} {478}},\ \bibinfo
  {pages} {926} (\bibinfo {year} {2018})},\ \Eprint
  {http://arxiv.org/abs/1702.05866} {arXiv:1702.05866 [astro-ph.HE]}
  \BibitemShut {NoStop}%
\bibitem [{\citenamefont {{Bykov}}\ \emph {et~al.}(2020)\citenamefont
  {{Bykov}}, \citenamefont {{Marcowith}}, \citenamefont {{Amato}},
  \citenamefont {{Kalyashova}}, \citenamefont {{Kruijssen}},\ and\
  \citenamefont {{Waxman}}}]{2020SSRv..216...42B}%
  \BibitemOpen
  \bibfield  {author} {\bibinfo {author} {\bibfnamefont {A.~M.}\ \bibnamefont
  {{Bykov}}}, \bibinfo {author} {\bibfnamefont {A.}~\bibnamefont
  {{Marcowith}}}, \bibinfo {author} {\bibfnamefont {E.}~\bibnamefont
  {{Amato}}}, \bibinfo {author} {\bibfnamefont {M.~E.}\ \bibnamefont
  {{Kalyashova}}}, \bibinfo {author} {\bibfnamefont {J.~M.~D.}\ \bibnamefont
  {{Kruijssen}}}, \ and\ \bibinfo {author} {\bibfnamefont {E.}~\bibnamefont
  {{Waxman}}},\ }\href {\doibase 10.1007/s11214-020-00663-0} {\bibfield
  {journal} {\bibinfo  {journal} {Space Sci. Rev.}\ }\textbf {\bibinfo {volume}
  {216}},\ \bibinfo {eid} {42} (\bibinfo {year} {2020})},\ \Eprint
  {http://arxiv.org/abs/2003.11534} {arXiv:2003.11534 [astro-ph.HE]}
  \BibitemShut {NoStop}%
\bibitem [{\citenamefont {{Sveshnikova, L. G.}}(2003)}]{Sveshnikova2003}%
  \BibitemOpen
  \bibfield  {author} {\bibinfo {author} {\bibnamefont {{Sveshnikova, L.
  G.}}},\ }\href {\doibase 10.1051/0004-6361:20030909} {\bibfield  {journal}
  {\bibinfo  {journal} {Astron.Astrophys.}\ }\textbf {\bibinfo {volume}
  {409}},\ \bibinfo {pages} {799} (\bibinfo {year} {2003})}\BibitemShut
  {NoStop}%
\bibitem [{\citenamefont {Murase}\ \emph {et~al.}(2014)\citenamefont {Murase},
  \citenamefont {Thompson},\ and\ \citenamefont {Ofek}}]{Murase:2013kda}%
  \BibitemOpen
  \bibfield  {author} {\bibinfo {author} {\bibfnamefont {K.}~\bibnamefont
  {Murase}}, \bibinfo {author} {\bibfnamefont {T.~A.}\ \bibnamefont
  {Thompson}}, \ and\ \bibinfo {author} {\bibfnamefont {E.~O.}\ \bibnamefont
  {Ofek}},\ }\href {\doibase 10.1093/mnras/stu384} {\bibfield  {journal}
  {\bibinfo  {journal} {Mon. Not. Roy. Astron. Soc.}\ }\textbf {\bibinfo
  {volume} {440}},\ \bibinfo {pages} {2528} (\bibinfo {year} {2014})},\ \Eprint
  {http://arxiv.org/abs/1311.6778} {arXiv:1311.6778 [astro-ph.HE]} \BibitemShut
  {NoStop}%
\bibitem [{\citenamefont {Inoue}\ \emph {et~al.}(2021)\citenamefont {Inoue},
  \citenamefont {Marcowith}, \citenamefont {Giacinti}, \citenamefont {van
  Marle},\ and\ \citenamefont {Nishino}}]{Inoue:2021bjx}%
  \BibitemOpen
  \bibfield  {author} {\bibinfo {author} {\bibfnamefont {T.}~\bibnamefont
  {Inoue}}, \bibinfo {author} {\bibfnamefont {A.}~\bibnamefont {Marcowith}},
  \bibinfo {author} {\bibfnamefont {G.}~\bibnamefont {Giacinti}}, \bibinfo
  {author} {\bibfnamefont {A.~J.}\ \bibnamefont {van Marle}}, \ and\ \bibinfo
  {author} {\bibfnamefont {S.}~\bibnamefont {Nishino}},\ }\href {\doibase
  10.3847/1538-4357/ac21ce} {\bibfield  {journal} {\bibinfo  {journal}
  {Astrophys. J.}\ }\textbf {\bibinfo {volume} {922}},\ \bibinfo {pages} {7}
  (\bibinfo {year} {2021})},\ \Eprint {http://arxiv.org/abs/2108.13433}
  {arXiv:2108.13433 [astro-ph.HE]} \BibitemShut {NoStop}%
\bibitem [{\citenamefont {Ioka}\ \emph {et~al.}(2017)\citenamefont {Ioka},
  \citenamefont {Matsumoto}, \citenamefont {Teraki}, \citenamefont
  {Kashiyama},\ and\ \citenamefont {Murase}}]{Ioka:2016bil}%
  \BibitemOpen
  \bibfield  {author} {\bibinfo {author} {\bibfnamefont {K.}~\bibnamefont
  {Ioka}}, \bibinfo {author} {\bibfnamefont {T.}~\bibnamefont {Matsumoto}},
  \bibinfo {author} {\bibfnamefont {Y.}~\bibnamefont {Teraki}}, \bibinfo
  {author} {\bibfnamefont {K.}~\bibnamefont {Kashiyama}}, \ and\ \bibinfo
  {author} {\bibfnamefont {K.}~\bibnamefont {Murase}},\ }\href {\doibase
  10.1093/mnras/stx1337} {\bibfield  {journal} {\bibinfo  {journal} {Mon. Not.
  Roy. Astron. Soc.}\ }\textbf {\bibinfo {volume} {470}},\ \bibinfo {pages}
  {3332} (\bibinfo {year} {2017})},\ \Eprint {http://arxiv.org/abs/1612.03913}
  {arXiv:1612.03913 [astro-ph.HE]} \BibitemShut {NoStop}%
\bibitem [{\citenamefont {Kimura}\ \emph {et~al.}(2025)\citenamefont {Kimura},
  \citenamefont {Tomida}, \citenamefont {Kobayashi}, \citenamefont {Kin},\ and\
  \citenamefont {Zhang}}]{Kimura:2024izc}%
  \BibitemOpen
  \bibfield  {author} {\bibinfo {author} {\bibfnamefont {S.~S.}\ \bibnamefont
  {Kimura}}, \bibinfo {author} {\bibfnamefont {K.}~\bibnamefont {Tomida}},
  \bibinfo {author} {\bibfnamefont {M.~I.~N.}\ \bibnamefont {Kobayashi}},
  \bibinfo {author} {\bibfnamefont {K.}~\bibnamefont {Kin}}, \ and\ \bibinfo
  {author} {\bibfnamefont {B.}~\bibnamefont {Zhang}},\ }\href {\doibase
  10.3847/2041-8213/adb841} {\bibfield  {journal} {\bibinfo  {journal}
  {Astrophys. J. Lett.}\ }\textbf {\bibinfo {volume} {981}},\ \bibinfo {pages}
  {L36} (\bibinfo {year} {2025})},\ \Eprint {http://arxiv.org/abs/2412.08136}
  {arXiv:2412.08136 [astro-ph.HE]} \BibitemShut {NoStop}%
\bibitem [{\citenamefont {Cao}\ \emph {et~al.}(2024{\natexlab{b}})\citenamefont
  {Cao} \emph {et~al.}}]{LHAASO:2024psv}%
  \BibitemOpen
  \bibfield  {author} {\bibinfo {author} {\bibfnamefont {Z.}~\bibnamefont
  {Cao}} \emph {et~al.} (\bibinfo {collaboration} {LHAASO}),\ }\href@noop {} {\
   (\bibinfo {year} {2024}{\natexlab{b}})},\ \Eprint
  {http://arxiv.org/abs/2410.08988} {arXiv:2410.08988 [astro-ph.HE]}
  \BibitemShut {NoStop}%
\bibitem [{\citenamefont {Mirabel}\ and\ \citenamefont
  {Rodriguez}(1994)}]{Mirabel:1994rb}%
  \BibitemOpen
  \bibfield  {author} {\bibinfo {author} {\bibfnamefont {I.~F.}\ \bibnamefont
  {Mirabel}}\ and\ \bibinfo {author} {\bibfnamefont {L.~F.}\ \bibnamefont
  {Rodriguez}},\ }\href {\doibase 10.1038/371046a0} {\bibfield  {journal}
  {\bibinfo  {journal} {Nature}\ }\textbf {\bibinfo {volume} {371}},\ \bibinfo
  {pages} {46} (\bibinfo {year} {1994})}\BibitemShut {NoStop}%
\bibitem [{\citenamefont {Kimura}\ \emph
  {et~al.}(2018{\natexlab{a}})\citenamefont {Kimura}, \citenamefont {Murase},\
  and\ \citenamefont {Zhang}}]{Kimura:2017ubz}%
  \BibitemOpen
  \bibfield  {author} {\bibinfo {author} {\bibfnamefont {S.~S.}\ \bibnamefont
  {Kimura}}, \bibinfo {author} {\bibfnamefont {K.}~\bibnamefont {Murase}}, \
  and\ \bibinfo {author} {\bibfnamefont {B.~T.}\ \bibnamefont {Zhang}},\ }\href
  {\doibase 10.1103/PhysRevD.97.023026} {\bibfield  {journal} {\bibinfo
  {journal} {Phys. Rev. D}\ }\textbf {\bibinfo {volume} {97}},\ \bibinfo
  {pages} {023026} (\bibinfo {year} {2018}{\natexlab{a}})},\ \Eprint
  {http://arxiv.org/abs/1705.05027} {arXiv:1705.05027 [astro-ph.HE]}
  \BibitemShut {NoStop}%
\bibitem [{\citenamefont {Caprioli}(2015)}]{Caprioli:2015zka}%
  \BibitemOpen
  \bibfield  {author} {\bibinfo {author} {\bibfnamefont {D.}~\bibnamefont
  {Caprioli}},\ }\href {\doibase 10.1088/2041-8205/811/2/L38} {\bibfield
  {journal} {\bibinfo  {journal} {Astrophys. J.}\ }\textbf {\bibinfo {volume}
  {811}},\ \bibinfo {pages} {L38} (\bibinfo {year} {2015})},\ \Eprint
  {http://arxiv.org/abs/1505.06739} {arXiv:1505.06739 [astro-ph.HE]}
  \BibitemShut {NoStop}%
\bibitem [{\citenamefont {Mbarek}\ and\ \citenamefont
  {Caprioli}(2019)}]{Mbarek:2019glq}%
  \BibitemOpen
  \bibfield  {author} {\bibinfo {author} {\bibfnamefont {R.}~\bibnamefont
  {Mbarek}}\ and\ \bibinfo {author} {\bibfnamefont {D.}~\bibnamefont
  {Caprioli}},\ }\href {\doibase 10.3847/1538-4357/ab4a08} {\bibfield
  {journal} {\bibinfo  {journal} {Astrophys. J.}\ }\textbf {\bibinfo {volume}
  {886}},\ \bibinfo {pages} {8} (\bibinfo {year} {2019})},\ \Eprint
  {http://arxiv.org/abs/1904.02720} {arXiv:1904.02720 [astro-ph.HE]}
  \BibitemShut {NoStop}%
\bibitem [{\citenamefont {Mbarek}\ \emph {et~al.}(2023)\citenamefont {Mbarek},
  \citenamefont {Caprioli},\ and\ \citenamefont {Murase}}]{Mbarek:2022nat}%
  \BibitemOpen
  \bibfield  {author} {\bibinfo {author} {\bibfnamefont {R.}~\bibnamefont
  {Mbarek}}, \bibinfo {author} {\bibfnamefont {D.}~\bibnamefont {Caprioli}}, \
  and\ \bibinfo {author} {\bibfnamefont {K.}~\bibnamefont {Murase}},\ }\href
  {\doibase 10.3847/1538-4357/aca481} {\bibfield  {journal} {\bibinfo
  {journal} {Astrophys. J.}\ }\textbf {\bibinfo {volume} {942}},\ \bibinfo
  {pages} {37} (\bibinfo {year} {2023})},\ \Eprint
  {http://arxiv.org/abs/2207.07130} {arXiv:2207.07130 [astro-ph.HE]}
  \BibitemShut {NoStop}%
\bibitem [{\citenamefont {Mbarek}\ \emph {et~al.}(2025)\citenamefont {Mbarek},
  \citenamefont {Caprioli},\ and\ \citenamefont {Murase}}]{Mbarek:2024nvv}%
  \BibitemOpen
  \bibfield  {author} {\bibinfo {author} {\bibfnamefont {R.}~\bibnamefont
  {Mbarek}}, \bibinfo {author} {\bibfnamefont {D.}~\bibnamefont {Caprioli}}, \
  and\ \bibinfo {author} {\bibfnamefont {K.}~\bibnamefont {Murase}},\ }\href
  {\doibase 10.1103/PhysRevD.111.023024} {\bibfield  {journal} {\bibinfo
  {journal} {Phys. Rev. D}\ }\textbf {\bibinfo {volume} {111}},\ \bibinfo
  {pages} {023024} (\bibinfo {year} {2025})},\ \Eprint
  {http://arxiv.org/abs/2410.05696} {arXiv:2410.05696 [astro-ph.HE]}
  \BibitemShut {NoStop}%
\bibitem [{\citenamefont {Cooper}\ \emph {et~al.}(2020)\citenamefont {Cooper},
  \citenamefont {Gaggero}, \citenamefont {Markoff},\ and\ \citenamefont
  {Zhang}}]{Cooper:2020tzq}%
  \BibitemOpen
  \bibfield  {author} {\bibinfo {author} {\bibfnamefont {A.~J.}\ \bibnamefont
  {Cooper}}, \bibinfo {author} {\bibfnamefont {D.}~\bibnamefont {Gaggero}},
  \bibinfo {author} {\bibfnamefont {S.}~\bibnamefont {Markoff}}, \ and\
  \bibinfo {author} {\bibfnamefont {S.}~\bibnamefont {Zhang}},\ }\href
  {\doibase 10.1093/mnras/staa373} {\bibfield  {journal} {\bibinfo  {journal}
  {Mon. Not. Roy. Astron. Soc.}\ }\textbf {\bibinfo {volume} {493}},\ \bibinfo
  {pages} {3212} (\bibinfo {year} {2020})},\ \Eprint
  {http://arxiv.org/abs/2002.01477} {arXiv:2002.01477 [astro-ph.HE]}
  \BibitemShut {NoStop}%
\bibitem [{\citenamefont {Kimura}\ \emph {et~al.}(2021)\citenamefont {Kimura},
  \citenamefont {Sudoh}, \citenamefont {Kashiyama},\ and\ \citenamefont
  {Kawanaka}}]{Kimura:2021eik}%
  \BibitemOpen
  \bibfield  {author} {\bibinfo {author} {\bibfnamefont {S.~S.}\ \bibnamefont
  {Kimura}}, \bibinfo {author} {\bibfnamefont {T.}~\bibnamefont {Sudoh}},
  \bibinfo {author} {\bibfnamefont {K.}~\bibnamefont {Kashiyama}}, \ and\
  \bibinfo {author} {\bibfnamefont {N.}~\bibnamefont {Kawanaka}},\ }\href
  {\doibase 10.3847/1538-4357/abff58} {\bibfield  {journal} {\bibinfo
  {journal} {Astrophys. J.}\ }\textbf {\bibinfo {volume} {915}},\ \bibinfo
  {pages} {31} (\bibinfo {year} {2021})},\ \Eprint
  {http://arxiv.org/abs/2103.15029} {arXiv:2103.15029 [astro-ph.HE]}
  \BibitemShut {NoStop}%
\bibitem [{\citenamefont {Yue}\ \emph {et~al.}(2024)\citenamefont {Yue},
  \citenamefont {Zhang}, \citenamefont {Ge}, \citenamefont {Nie}, \citenamefont
  {Zhang}, \citenamefont {Liu}, \citenamefont {Guo},\ and\ \citenamefont
  {Hu}}]{Yue:2024nns}%
  \BibitemOpen
  \bibfield  {author} {\bibinfo {author} {\bibfnamefont {H.}~\bibnamefont
  {Yue}}, \bibinfo {author} {\bibfnamefont {J.}~\bibnamefont {Zhang}}, \bibinfo
  {author} {\bibfnamefont {Y.}~\bibnamefont {Ge}}, \bibinfo {author}
  {\bibfnamefont {L.}~\bibnamefont {Nie}}, \bibinfo {author} {\bibfnamefont
  {P.}~\bibnamefont {Zhang}}, \bibinfo {author} {\bibfnamefont
  {W.}~\bibnamefont {Liu}}, \bibinfo {author} {\bibfnamefont {Y.}~\bibnamefont
  {Guo}}, \ and\ \bibinfo {author} {\bibfnamefont {H.}~\bibnamefont {Hu}},\
  }\href@noop {} {\  (\bibinfo {year} {2024})},\ \Eprint
  {http://arxiv.org/abs/2412.13889} {arXiv:2412.13889 [astro-ph.HE]}
  \BibitemShut {NoStop}%
\bibitem [{\citenamefont {{Tetarenko}}\ \emph {et~al.}(2016)\citenamefont
  {{Tetarenko}}, \citenamefont {{Sivakoff}}, \citenamefont {{Heinke}},\ and\
  \citenamefont {{Gladstone}}}]{tetarenko:2016aa}%
  \BibitemOpen
  \bibfield  {author} {\bibinfo {author} {\bibfnamefont {B.~E.}\ \bibnamefont
  {{Tetarenko}}}, \bibinfo {author} {\bibfnamefont {G.~R.}\ \bibnamefont
  {{Sivakoff}}}, \bibinfo {author} {\bibfnamefont {C.~O.}\ \bibnamefont
  {{Heinke}}}, \ and\ \bibinfo {author} {\bibfnamefont {J.~C.}\ \bibnamefont
  {{Gladstone}}},\ }\href {\doibase 10.3847/0067-0049/222/2/15} {\bibfield
  {journal} {\bibinfo  {journal} {Astrophys.J.Suppl.}\ }\textbf {\bibinfo
  {volume} {222}},\ \bibinfo {eid} {15} (\bibinfo {year} {2016})},\ \Eprint
  {http://arxiv.org/abs/1512.00778} {arXiv:1512.00778 [astro-ph.HE]}
  \BibitemShut {NoStop}%
\bibitem [{\citenamefont {{Corral-Santana}}\ \emph {et~al.}(2016)\citenamefont
  {{Corral-Santana}}, \citenamefont {{Casares}}, \citenamefont
  {{Mu{\~n}oz-Darias}}, \citenamefont {{Bauer}}, \citenamefont
  {{Mart{\'\i}nez-Pais}},\ and\ \citenamefont
  {{Russell}}}]{Corral-Santana:2016aa}%
  \BibitemOpen
  \bibfield  {author} {\bibinfo {author} {\bibfnamefont {J.~M.}\ \bibnamefont
  {{Corral-Santana}}}, \bibinfo {author} {\bibfnamefont {J.}~\bibnamefont
  {{Casares}}}, \bibinfo {author} {\bibfnamefont {T.}~\bibnamefont
  {{Mu{\~n}oz-Darias}}}, \bibinfo {author} {\bibfnamefont {F.~E.}\ \bibnamefont
  {{Bauer}}}, \bibinfo {author} {\bibfnamefont {I.~G.}\ \bibnamefont
  {{Mart{\'\i}nez-Pais}}}, \ and\ \bibinfo {author} {\bibfnamefont {D.~M.}\
  \bibnamefont {{Russell}}},\ }\href {\doibase 10.1051/0004-6361/201527130}
  {\bibfield  {journal} {\bibinfo  {journal} {Astron.Astrophys.}\ }\textbf
  {\bibinfo {volume} {587}},\ \bibinfo {eid} {A61} (\bibinfo {year} {2016})},\
  \Eprint {http://arxiv.org/abs/1510.08869} {arXiv:1510.08869 [astro-ph.HE]}
  \BibitemShut {NoStop}%
\bibitem [{\citenamefont {{Done}}\ \emph {et~al.}(2007)\citenamefont {{Done}},
  \citenamefont {{Gierli{\'n}ski}},\ and\ \citenamefont
  {{Kubota}}}]{Done2007aa}%
  \BibitemOpen
  \bibfield  {author} {\bibinfo {author} {\bibfnamefont {C.}~\bibnamefont
  {{Done}}}, \bibinfo {author} {\bibfnamefont {M.}~\bibnamefont
  {{Gierli{\'n}ski}}}, \ and\ \bibinfo {author} {\bibfnamefont
  {A.}~\bibnamefont {{Kubota}}},\ }\href {\doibase 10.1007/s00159-007-0006-1}
  {\bibfield  {journal} {\bibinfo  {journal} {Astron. Astrophys. Rev.}\
  }\textbf {\bibinfo {volume} {15}},\ \bibinfo {pages} {1} (\bibinfo {year}
  {2007})},\ \Eprint {http://arxiv.org/abs/0708.0148} {arXiv:0708.0148
  [astro-ph]} \BibitemShut {NoStop}%
\bibitem [{\citenamefont {Alfaro}\ \emph {et~al.}(2017)\citenamefont {Alfaro}
  \emph {et~al.}}]{HAWC:2017osk}%
  \BibitemOpen
  \bibfield  {author} {\bibinfo {author} {\bibfnamefont {R.}~\bibnamefont
  {Alfaro}} \emph {et~al.} (\bibinfo {collaboration} {HAWC}),\ }\href {\doibase
  10.1103/PhysRevD.96.122001} {\bibfield  {journal} {\bibinfo  {journal} {Phys.
  Rev. D}\ }\textbf {\bibinfo {volume} {96}},\ \bibinfo {pages} {122001}
  (\bibinfo {year} {2017})},\ \Eprint {http://arxiv.org/abs/1710.00890}
  {arXiv:1710.00890 [astro-ph.HE]} \BibitemShut {NoStop}%
\bibitem [{\citenamefont {Aharonian}\ \emph {et~al.}(2024)\citenamefont
  {Aharonian} \emph {et~al.}}]{HESS:2024rlh}%
  \BibitemOpen
  \bibfield  {author} {\bibinfo {author} {\bibfnamefont {F.}~\bibnamefont
  {Aharonian}} \emph {et~al.} (\bibinfo {collaboration} {H.E.S.S.}),\ }\href
  {\doibase 10.1126/science.adi2048} {\bibfield  {journal} {\bibinfo  {journal}
  {Science}\ }\textbf {\bibinfo {volume} {383}},\ \bibinfo {pages} {adi2048}
  (\bibinfo {year} {2024})},\ \Eprint {http://arxiv.org/abs/2401.16019}
  {arXiv:2401.16019 [astro-ph.HE]} \BibitemShut {NoStop}%
\bibitem [{\citenamefont {Fang}\ \emph {et~al.}(2024)\citenamefont {Fang},
  \citenamefont {Halzen}, \citenamefont {Heinz},\ and\ \citenamefont
  {Gallagher}}]{Fang:2024wmf}%
  \BibitemOpen
  \bibfield  {author} {\bibinfo {author} {\bibfnamefont {K.}~\bibnamefont
  {Fang}}, \bibinfo {author} {\bibfnamefont {F.}~\bibnamefont {Halzen}},
  \bibinfo {author} {\bibfnamefont {S.}~\bibnamefont {Heinz}}, \ and\ \bibinfo
  {author} {\bibfnamefont {J.~S.}\ \bibnamefont {Gallagher}},\ }\href {\doibase
  10.3847/2041-8213/ad887b} {\bibfield  {journal} {\bibinfo  {journal}
  {Astrophys. J. Lett.}\ }\textbf {\bibinfo {volume} {975}},\ \bibinfo {pages}
  {L35} (\bibinfo {year} {2024})},\ \Eprint {http://arxiv.org/abs/2410.02119}
  {arXiv:2410.02119 [astro-ph.HE]} \BibitemShut {NoStop}%
\bibitem [{\citenamefont {Kuze}\ \emph {et~al.}(2025)\citenamefont {Kuze},
  \citenamefont {Kimura},\ and\ \citenamefont {Fang}}]{Kuze:2025wda}%
  \BibitemOpen
  \bibfield  {author} {\bibinfo {author} {\bibfnamefont {R.}~\bibnamefont
  {Kuze}}, \bibinfo {author} {\bibfnamefont {S.~S.}\ \bibnamefont {Kimura}}, \
  and\ \bibinfo {author} {\bibfnamefont {K.}~\bibnamefont {Fang}},\ }\href
  {\doibase 10.3847/1538-4357/adcc1b} {\bibfield  {journal} {\bibinfo
  {journal} {Astrophys. J.}\ }\textbf {\bibinfo {volume} {985}},\ \bibinfo
  {pages} {139} (\bibinfo {year} {2025})},\ \Eprint
  {http://arxiv.org/abs/2501.17467} {arXiv:2501.17467 [astro-ph.HE]}
  \BibitemShut {NoStop}%
\bibitem [{\citenamefont {{Abell}}\ and\ \citenamefont
  {{Margon}}(1979)}]{Abell:1979aa}%
  \BibitemOpen
  \bibfield  {author} {\bibinfo {author} {\bibfnamefont {G.~O.}\ \bibnamefont
  {{Abell}}}\ and\ \bibinfo {author} {\bibfnamefont {B.}~\bibnamefont
  {{Margon}}},\ }\href {\doibase 10.1038/279701a0} {\bibfield  {journal}
  {\bibinfo  {journal} {\nat}\ }\textbf {\bibinfo {volume} {279}},\ \bibinfo
  {pages} {701} (\bibinfo {year} {1979})}\BibitemShut {NoStop}%
\bibitem [{\citenamefont {{Safi-Harb}}\ and\ \citenamefont
  {{{\"O}gelman}}(1997)}]{Safi-harv:1997aa}%
  \BibitemOpen
  \bibfield  {author} {\bibinfo {author} {\bibfnamefont {S.}~\bibnamefont
  {{Safi-Harb}}}\ and\ \bibinfo {author} {\bibfnamefont {H.}~\bibnamefont
  {{{\"O}gelman}}},\ }\href {\doibase 10.1086/304274} {\bibfield  {journal}
  {\bibinfo  {journal} {\apj}\ }\textbf {\bibinfo {volume} {483}},\ \bibinfo
  {pages} {868} (\bibinfo {year} {1997})}\BibitemShut {NoStop}%
\bibitem [{\citenamefont {{Dubner}}\ \emph {et~al.}(1998)\citenamefont
  {{Dubner}}, \citenamefont {{Holdaway}}, \citenamefont {{Goss}},\ and\
  \citenamefont {{Mirabel}}}]{Dubner:1998aa}%
  \BibitemOpen
  \bibfield  {author} {\bibinfo {author} {\bibfnamefont {G.~M.}\ \bibnamefont
  {{Dubner}}}, \bibinfo {author} {\bibfnamefont {M.}~\bibnamefont
  {{Holdaway}}}, \bibinfo {author} {\bibfnamefont {W.~M.}\ \bibnamefont
  {{Goss}}}, \ and\ \bibinfo {author} {\bibfnamefont {I.~F.}\ \bibnamefont
  {{Mirabel}}},\ }\href {\doibase 10.1086/300537} {\bibfield  {journal}
  {\bibinfo  {journal} {Astron.J.}\ }\textbf {\bibinfo {volume} {116}},\
  \bibinfo {pages} {1842} (\bibinfo {year} {1998})}\BibitemShut {NoStop}%
\bibitem [{\citenamefont {{Sakemi}}\ \emph {et~al.}(2023)\citenamefont
  {{Sakemi}}, \citenamefont {{Machida}}, \citenamefont {{Yamamoto}},\ and\
  \citenamefont {{Tachihara}}}]{Sakemi:2023aa}%
  \BibitemOpen
  \bibfield  {author} {\bibinfo {author} {\bibfnamefont {H.}~\bibnamefont
  {{Sakemi}}}, \bibinfo {author} {\bibfnamefont {M.}~\bibnamefont {{Machida}}},
  \bibinfo {author} {\bibfnamefont {H.}~\bibnamefont {{Yamamoto}}}, \ and\
  \bibinfo {author} {\bibfnamefont {K.}~\bibnamefont {{Tachihara}}},\ }\href
  {\doibase 10.1093/pasj/psad001} {\bibfield  {journal} {\bibinfo  {journal}
  {Publ.Astron.Soc.Jap.}\ }\textbf {\bibinfo {volume} {75}},\ \bibinfo {pages}
  {338} (\bibinfo {year} {2023})},\ \Eprint {http://arxiv.org/abs/2301.13333}
  {arXiv:2301.13333 [astro-ph.GA]} \BibitemShut {NoStop}%
\bibitem [{\citenamefont {Abeysekara}\ \emph {et~al.}(2018)\citenamefont
  {Abeysekara} \emph {et~al.}}]{HAWC:2018gwz}%
  \BibitemOpen
  \bibfield  {author} {\bibinfo {author} {\bibfnamefont {A.~U.}\ \bibnamefont
  {Abeysekara}} \emph {et~al.} (\bibinfo {collaboration} {HAWC}),\ }\href
  {\doibase 10.1038/s41586-018-0565-5} {\bibfield  {journal} {\bibinfo
  {journal} {Nature}\ }\textbf {\bibinfo {volume} {562}},\ \bibinfo {pages}
  {82} (\bibinfo {year} {2018})},\ \bibinfo {note} {[Erratum: Nature 564, E38
  (2018)]},\ \Eprint {http://arxiv.org/abs/1810.01892} {arXiv:1810.01892
  [astro-ph.HE]} \BibitemShut {NoStop}%
\bibitem [{\citenamefont {Alfaro}\ \emph
  {et~al.}(2024{\natexlab{a}})\citenamefont {Alfaro} \emph
  {et~al.}}]{HAWC:2024ysp}%
  \BibitemOpen
  \bibfield  {author} {\bibinfo {author} {\bibfnamefont {R.}~\bibnamefont
  {Alfaro}} \emph {et~al.} (\bibinfo {collaboration} {HAWC}),\ }\href {\doibase
  10.3847/1538-4357/ad7e1b} {\bibfield  {journal} {\bibinfo  {journal}
  {Astrophys. J.}\ }\textbf {\bibinfo {volume} {976}},\ \bibinfo {pages} {30}
  (\bibinfo {year} {2024}{\natexlab{a}})},\ \Eprint
  {http://arxiv.org/abs/2410.21796} {arXiv:2410.21796 [astro-ph.HE]}
  \BibitemShut {NoStop}%
\bibitem [{Note1()}]{Note1}%
  \BibitemOpen
  \bibinfo {note} {In Table~1 of Ref.~\cite {Kimura:2020aa}, $B$ is given in
  the unit of $\mu $G not G.}\BibitemShut {Stop}%
\bibitem [{\citenamefont {{Sudoh}}\ \emph {et~al.}(2020)\citenamefont
  {{Sudoh}}, \citenamefont {{Inoue}},\ and\ \citenamefont
  {{Khangulyan}}}]{Sudoh:2020aa}%
  \BibitemOpen
  \bibfield  {author} {\bibinfo {author} {\bibfnamefont {T.}~\bibnamefont
  {{Sudoh}}}, \bibinfo {author} {\bibfnamefont {Y.}~\bibnamefont {{Inoue}}}, \
  and\ \bibinfo {author} {\bibfnamefont {D.}~\bibnamefont {{Khangulyan}}},\
  }\href {\doibase 10.3847/1538-4357/ab6442} {\bibfield  {journal} {\bibinfo
  {journal} {\apj}\ }\textbf {\bibinfo {volume} {889}},\ \bibinfo {eid} {146}
  (\bibinfo {year} {2020})},\ \Eprint {http://arxiv.org/abs/1911.00013}
  {arXiv:1911.00013 [astro-ph.HE]} \BibitemShut {NoStop}%
\bibitem [{\citenamefont {{Kimura}}\ \emph {et~al.}(2020)\citenamefont
  {{Kimura}}, \citenamefont {{Murase}},\ and\ \citenamefont
  {{M{\'e}sz{\'a}ros}}}]{Kimura:2020aa}%
  \BibitemOpen
  \bibfield  {author} {\bibinfo {author} {\bibfnamefont {S.~S.}\ \bibnamefont
  {{Kimura}}}, \bibinfo {author} {\bibfnamefont {K.}~\bibnamefont {{Murase}}},
  \ and\ \bibinfo {author} {\bibfnamefont {P.}~\bibnamefont
  {{M{\'e}sz{\'a}ros}}},\ }\href {\doibase 10.3847/1538-4357/abbe00} {\bibfield
   {journal} {\bibinfo  {journal} {\apj}\ }\textbf {\bibinfo {volume} {904}},\
  \bibinfo {eid} {188} (\bibinfo {year} {2020})},\ \Eprint
  {http://arxiv.org/abs/2008.04515} {arXiv:2008.04515 [astro-ph.HE]}
  \BibitemShut {NoStop}%
\bibitem [{\citenamefont {Ohmura}\ \emph {et~al.}(2021)\citenamefont {Ohmura},
  \citenamefont {Ono}, \citenamefont {Sakemi}, \citenamefont {Tashima},
  \citenamefont {Omae},\ and\ \citenamefont {Machida}}]{Ohmura:2021skq}%
  \BibitemOpen
  \bibfield  {author} {\bibinfo {author} {\bibfnamefont {T.}~\bibnamefont
  {Ohmura}}, \bibinfo {author} {\bibfnamefont {K.}~\bibnamefont {Ono}},
  \bibinfo {author} {\bibfnamefont {H.}~\bibnamefont {Sakemi}}, \bibinfo
  {author} {\bibfnamefont {Y.}~\bibnamefont {Tashima}}, \bibinfo {author}
  {\bibfnamefont {R.}~\bibnamefont {Omae}}, \ and\ \bibinfo {author}
  {\bibfnamefont {M.}~\bibnamefont {Machida}},\ }\href {\doibase
  10.3847/1538-4357/abe5a1} {\bibfield  {journal} {\bibinfo  {journal}
  {Astrophys. J.}\ }\textbf {\bibinfo {volume} {910}},\ \bibinfo {pages} {149}
  (\bibinfo {year} {2021})},\ \Eprint {http://arxiv.org/abs/2102.06728}
  {arXiv:2102.06728 [astro-ph.HE]} \BibitemShut {NoStop}%
\bibitem [{\citenamefont {Alfaro}\ \emph
  {et~al.}(2024{\natexlab{b}})\citenamefont {Alfaro} \emph
  {et~al.}}]{Alfaro:2024cjd}%
  \BibitemOpen
  \bibfield  {author} {\bibinfo {author} {\bibfnamefont {R.}~\bibnamefont
  {Alfaro}} \emph {et~al.},\ }\href {\doibase 10.1038/s41586-024-07995-9}
  {\bibfield  {journal} {\bibinfo  {journal} {Nature}\ }\textbf {\bibinfo
  {volume} {634}},\ \bibinfo {pages} {557} (\bibinfo {year}
  {2024}{\natexlab{b}})},\ \Eprint {http://arxiv.org/abs/2410.16117}
  {arXiv:2410.16117 [astro-ph.HE]} \BibitemShut {NoStop}%
\bibitem [{\citenamefont {Suzuki}\ \emph {et~al.}(2025)\citenamefont {Suzuki}
  \emph {et~al.}}]{Suzuki:2024rzc}%
  \BibitemOpen
  \bibfield  {author} {\bibinfo {author} {\bibfnamefont {H.}~\bibnamefont
  {Suzuki}} \emph {et~al.},\ }\href {\doibase 10.3847/2041-8213/ad9d11}
  {\bibfield  {journal} {\bibinfo  {journal} {Astrophys. J. Lett.}\ }\textbf
  {\bibinfo {volume} {978}},\ \bibinfo {pages} {L20} (\bibinfo {year}
  {2025})},\ \Eprint {http://arxiv.org/abs/2412.08089} {arXiv:2412.08089
  [astro-ph.HE]} \BibitemShut {NoStop}%
\bibitem [{\citenamefont {Fender}\ \emph {et~al.}(1999)\citenamefont {Fender},
  \citenamefont {Garrington}, \citenamefont {McKay}, \citenamefont {Muxlow},
  \citenamefont {Pooley}, \citenamefont {Spencer}, \citenamefont {Stirling},\
  and\ \citenamefont {Waltman}}]{Fender:1998nq}%
  \BibitemOpen
  \bibfield  {author} {\bibinfo {author} {\bibfnamefont {R.~P.}\ \bibnamefont
  {Fender}}, \bibinfo {author} {\bibfnamefont {S.~T.}\ \bibnamefont
  {Garrington}}, \bibinfo {author} {\bibfnamefont {D.~J.}\ \bibnamefont
  {McKay}}, \bibinfo {author} {\bibfnamefont {T.~W.~B.}\ \bibnamefont
  {Muxlow}}, \bibinfo {author} {\bibfnamefont {G.~G.}\ \bibnamefont {Pooley}},
  \bibinfo {author} {\bibfnamefont {R.~E.}\ \bibnamefont {Spencer}}, \bibinfo
  {author} {\bibfnamefont {A.~M.}\ \bibnamefont {Stirling}}, \ and\ \bibinfo
  {author} {\bibfnamefont {E.~B.}\ \bibnamefont {Waltman}},\ }\href {\doibase
  10.1046/j.1365-8711.1999.02364.x} {\bibfield  {journal} {\bibinfo  {journal}
  {Mon. Not. Roy. Astron. Soc.}\ }\textbf {\bibinfo {volume} {304}},\ \bibinfo
  {pages} {865} (\bibinfo {year} {1999})},\ \Eprint
  {http://arxiv.org/abs/astro-ph/9812150} {arXiv:astro-ph/9812150} \BibitemShut
  {NoStop}%
\bibitem [{\citenamefont {Done}\ \emph {et~al.}(2004)\citenamefont {Done},
  \citenamefont {Wardzinski},\ and\ \citenamefont {Gierlinski}}]{Done:2003sz}%
  \BibitemOpen
  \bibfield  {author} {\bibinfo {author} {\bibfnamefont {C.}~\bibnamefont
  {Done}}, \bibinfo {author} {\bibfnamefont {G.}~\bibnamefont {Wardzinski}}, \
  and\ \bibinfo {author} {\bibfnamefont {M.}~\bibnamefont {Gierlinski}},\
  }\href {\doibase 10.1111/j.1365-2966.2004.07545.x} {\bibfield  {journal}
  {\bibinfo  {journal} {Mon. Not. Roy. Astron. Soc.}\ }\textbf {\bibinfo
  {volume} {349}},\ \bibinfo {pages} {393} (\bibinfo {year} {2004})},\ \Eprint
  {http://arxiv.org/abs/astro-ph/0308536} {arXiv:astro-ph/0308536} \BibitemShut
  {NoStop}%
\bibitem [{\citenamefont {Koljonen}\ and\ \citenamefont
  {Hovatta}(2021)}]{Koljonen:2021dqn}%
  \BibitemOpen
  \bibfield  {author} {\bibinfo {author} {\bibfnamefont {K.~I.~I.}\
  \bibnamefont {Koljonen}}\ and\ \bibinfo {author} {\bibfnamefont
  {T.}~\bibnamefont {Hovatta}},\ }\href {\doibase 10.1051/0004-6361/202039581}
  {\bibfield  {journal} {\bibinfo  {journal} {Astron. Astrophys.}\ }\textbf
  {\bibinfo {volume} {647}},\ \bibinfo {pages} {A173} (\bibinfo {year}
  {2021})},\ \Eprint {http://arxiv.org/abs/2102.00693} {arXiv:2102.00693
  [astro-ph.HE]} \BibitemShut {NoStop}%
\bibitem [{\citenamefont {Tetarenko}\ \emph {et~al.}(2018)\citenamefont
  {Tetarenko}, \citenamefont {Freeman}, \citenamefont {Rosolowsky},
  \citenamefont {Miller-Jones},\ and\ \citenamefont
  {Sivakoff}}]{Tetarenko:2017zke}%
  \BibitemOpen
  \bibfield  {author} {\bibinfo {author} {\bibfnamefont {A.~J.}\ \bibnamefont
  {Tetarenko}}, \bibinfo {author} {\bibfnamefont {P.}~\bibnamefont {Freeman}},
  \bibinfo {author} {\bibfnamefont {E.~W.}\ \bibnamefont {Rosolowsky}},
  \bibinfo {author} {\bibfnamefont {J.~C.~A.}\ \bibnamefont {Miller-Jones}}, \
  and\ \bibinfo {author} {\bibfnamefont {G.~R.}\ \bibnamefont {Sivakoff}},\
  }\href {\doibase 10.1093/mnras/stx3151} {\bibfield  {journal} {\bibinfo
  {journal} {Mon. Not. Roy. Astron. Soc.}\ }\textbf {\bibinfo {volume} {475}},\
  \bibinfo {pages} {448} (\bibinfo {year} {2018})},\ \Eprint
  {http://arxiv.org/abs/1712.00432} {arXiv:1712.00432 [astro-ph.HE]}
  \BibitemShut {NoStop}%
\bibitem [{\citenamefont {Dopita}\ \emph {et~al.}(2012)\citenamefont {Dopita},
  \citenamefont {Payne}, \citenamefont {Filipovic},\ and\ \citenamefont
  {Pannuti}}]{Dopita:2012aa}%
  \BibitemOpen
  \bibfield  {author} {\bibinfo {author} {\bibfnamefont {M.~A.}\ \bibnamefont
  {Dopita}}, \bibinfo {author} {\bibfnamefont {J.~L.}\ \bibnamefont {Payne}},
  \bibinfo {author} {\bibfnamefont {M.~D.}\ \bibnamefont {Filipovic}}, \ and\
  \bibinfo {author} {\bibfnamefont {T.~G.}\ \bibnamefont {Pannuti}},\ }\href
  {\doibase 10.1111/j.1365-2966.2012.21947.x} {\bibfield  {journal} {\bibinfo
  {journal} {Mon. Not. Roy. Astron. Soc.}\ }\textbf {\bibinfo {volume} {427}},\
  \bibinfo {pages} {956} (\bibinfo {year} {2012})},\ \Eprint
  {http://arxiv.org/abs/1208.3500} {arXiv:1208.3500 [astro-ph.HE]} \BibitemShut
  {NoStop}%
\bibitem [{\citenamefont {Abaroa}\ \emph {et~al.}(2024)\citenamefont {Abaroa},
  \citenamefont {Romero}, \citenamefont {Mancuso},\ and\ \citenamefont
  {Rizzo}}]{Abaroa:2024vnw}%
  \BibitemOpen
  \bibfield  {author} {\bibinfo {author} {\bibfnamefont {L.}~\bibnamefont
  {Abaroa}}, \bibinfo {author} {\bibfnamefont {G.~E.}\ \bibnamefont {Romero}},
  \bibinfo {author} {\bibfnamefont {G.~C.}\ \bibnamefont {Mancuso}}, \ and\
  \bibinfo {author} {\bibfnamefont {F.~N.}\ \bibnamefont {Rizzo}},\ }\href
  {\doibase 10.1051/0004-6361/202450202} {\bibfield  {journal} {\bibinfo
  {journal} {Astron. Astrophys.}\ }\textbf {\bibinfo {volume} {691}},\ \bibinfo
  {pages} {A93} (\bibinfo {year} {2024})},\ \Eprint
  {http://arxiv.org/abs/2409.16315} {arXiv:2409.16315 [astro-ph.HE]}
  \BibitemShut {NoStop}%
\bibitem [{\citenamefont {Blandford}(2000)}]{Blandford:1999hi}%
  \BibitemOpen
  \bibfield  {author} {\bibinfo {author} {\bibfnamefont {R.~D.}\ \bibnamefont
  {Blandford}},\ }\href {\doibase 10.1238/Physica.Topical.085a00191} {\bibfield
   {journal} {\bibinfo  {journal} {Phys. Scripta T}\ }\textbf {\bibinfo
  {volume} {85}},\ \bibinfo {pages} {191} (\bibinfo {year} {2000})},\ \Eprint
  {http://arxiv.org/abs/astro-ph/9906026} {arXiv:astro-ph/9906026} \BibitemShut
  {NoStop}%
\bibitem [{\citenamefont {Hillas}(1984)}]{Hillas:1985is}%
  \BibitemOpen
  \bibfield  {author} {\bibinfo {author} {\bibfnamefont {A.~M.}\ \bibnamefont
  {Hillas}},\ }\href {\doibase 10.1146/annurev.aa.22.090184.002233} {\bibfield
  {journal} {\bibinfo  {journal} {Ann. Rev. Astron. Astrophys.}\ }\textbf
  {\bibinfo {volume} {22}},\ \bibinfo {pages} {425} (\bibinfo {year}
  {1984})}\BibitemShut {NoStop}%
\bibitem [{\citenamefont {{Berezhko}}\ and\ \citenamefont
  {{Krymskii}}(1981)}]{1981SvAL....7..352B}%
  \BibitemOpen
  \bibfield  {author} {\bibinfo {author} {\bibfnamefont {E.~G.}\ \bibnamefont
  {{Berezhko}}}\ and\ \bibinfo {author} {\bibfnamefont {G.~F.}\ \bibnamefont
  {{Krymskii}}},\ }\href@noop {} {\bibfield  {journal} {\bibinfo  {journal}
  {Soviet Astronomy Letters}\ }\textbf {\bibinfo {volume} {7}},\ \bibinfo
  {pages} {352} (\bibinfo {year} {1981})}\BibitemShut {NoStop}%
\bibitem [{\citenamefont {{Earl}}\ \emph {et~al.}(1988)\citenamefont {{Earl}},
  \citenamefont {{Jokipii}},\ and\ \citenamefont
  {{Morfill}}}]{1988ApJ...331L..91E}%
  \BibitemOpen
  \bibfield  {author} {\bibinfo {author} {\bibfnamefont {J.~A.}\ \bibnamefont
  {{Earl}}}, \bibinfo {author} {\bibfnamefont {J.~R.}\ \bibnamefont
  {{Jokipii}}}, \ and\ \bibinfo {author} {\bibfnamefont {G.}~\bibnamefont
  {{Morfill}}},\ }\href {\doibase 10.1086/185242} {\bibfield  {journal}
  {\bibinfo  {journal} {Astrophys. J. Lett.}\ }\textbf {\bibinfo {volume}
  {331}},\ \bibinfo {pages} {L91} (\bibinfo {year} {1988})}\BibitemShut
  {NoStop}%
\bibitem [{\citenamefont {{Ostrowski}}(1990)}]{1990A&A...238..435O}%
  \BibitemOpen
  \bibfield  {author} {\bibinfo {author} {\bibfnamefont {M.}~\bibnamefont
  {{Ostrowski}}},\ }\href@noop {} {\bibfield  {journal} {\bibinfo  {journal}
  {Astron.Astrophys.}\ }\textbf {\bibinfo {volume} {238}},\ \bibinfo {pages}
  {435} (\bibinfo {year} {1990})}\BibitemShut {NoStop}%
\bibitem [{\citenamefont {{Webb}}(1989)}]{1989ApJ...340.1112W}%
  \BibitemOpen
  \bibfield  {author} {\bibinfo {author} {\bibfnamefont {G.~M.}\ \bibnamefont
  {{Webb}}},\ }\href {\doibase 10.1086/167462} {\bibfield  {journal} {\bibinfo
  {journal} {Astrophys. J.}\ }\textbf {\bibinfo {volume} {340}},\ \bibinfo
  {pages} {1112} (\bibinfo {year} {1989})}\BibitemShut {NoStop}%
\bibitem [{\citenamefont {{Rieger}}\ and\ \citenamefont
  {{Duffy}}(2004)}]{Rieger2004}%
  \BibitemOpen
  \bibfield  {author} {\bibinfo {author} {\bibfnamefont {F.~M.}\ \bibnamefont
  {{Rieger}}}\ and\ \bibinfo {author} {\bibfnamefont {P.}~\bibnamefont
  {{Duffy}}},\ }\href {\doibase 10.1086/425167} {\bibfield  {journal} {\bibinfo
   {journal} {Astrophys. J.}\ }\textbf {\bibinfo {volume} {617}},\ \bibinfo
  {pages} {155} (\bibinfo {year} {2004})},\ \Eprint
  {http://arxiv.org/abs/astro-ph/0410269} {arXiv:astro-ph/0410269 [astro-ph]}
  \BibitemShut {NoStop}%
\bibitem [{\citenamefont {Rieger}(2019)}]{Rieger:2019uyp}%
  \BibitemOpen
  \bibfield  {author} {\bibinfo {author} {\bibfnamefont {F.~M.}\ \bibnamefont
  {Rieger}},\ }\href {\doibase 10.3390/galaxies7030078} {\bibfield  {journal}
  {\bibinfo  {journal} {Galaxies}\ }\textbf {\bibinfo {volume} {7}},\ \bibinfo
  {pages} {78} (\bibinfo {year} {2019})},\ \Eprint
  {http://arxiv.org/abs/1909.07237} {arXiv:1909.07237 [astro-ph.HE]}
  \BibitemShut {NoStop}%
\bibitem [{\citenamefont {{Matthews}}\ and\ \citenamefont
  {{Taylor}}(2021)}]{2021MNRAS.503.5948M}%
  \BibitemOpen
  \bibfield  {author} {\bibinfo {author} {\bibfnamefont {J.~H.}\ \bibnamefont
  {{Matthews}}}\ and\ \bibinfo {author} {\bibfnamefont {A.~M.}\ \bibnamefont
  {{Taylor}}},\ }\href {\doibase 10.1093/mnras/stab758} {\bibfield  {journal}
  {\bibinfo  {journal} {Mon. Not. Roy. Astron. Soc.}\ }\textbf {\bibinfo
  {volume} {503}},\ \bibinfo {pages} {5948} (\bibinfo {year} {2021})},\ \Eprint
  {http://arxiv.org/abs/2103.06900} {arXiv:2103.06900 [astro-ph.HE]}
  \BibitemShut {NoStop}%
\bibitem [{\citenamefont {Wan}\ \emph {et~al.}(2025)\citenamefont {Wan},
  \citenamefont {Wang},\ and\ \citenamefont {Liu}}]{Wan:2025eoh}%
  \BibitemOpen
  \bibfield  {author} {\bibinfo {author} {\bibfnamefont {S.-Y.}\ \bibnamefont
  {Wan}}, \bibinfo {author} {\bibfnamefont {J.-s.}\ \bibnamefont {Wang}}, \
  and\ \bibinfo {author} {\bibfnamefont {R.-Y.}\ \bibnamefont {Liu}},\
  }\href@noop {} {\  (\bibinfo {year} {2025})},\ \Eprint
  {http://arxiv.org/abs/2507.02763} {arXiv:2507.02763 [astro-ph.HE]}
  \BibitemShut {NoStop}%
\bibitem [{\citenamefont {Seo}\ \emph {et~al.}(2024)\citenamefont {Seo},
  \citenamefont {Ryu},\ and\ \citenamefont {Kang}}]{Seo:2023dkc}%
  \BibitemOpen
  \bibfield  {author} {\bibinfo {author} {\bibfnamefont {J.}~\bibnamefont
  {Seo}}, \bibinfo {author} {\bibfnamefont {D.}~\bibnamefont {Ryu}}, \ and\
  \bibinfo {author} {\bibfnamefont {H.}~\bibnamefont {Kang}},\ }\href {\doibase
  10.3847/1538-4357/ad182c} {\bibfield  {journal} {\bibinfo  {journal}
  {Astrophys. J.}\ }\textbf {\bibinfo {volume} {962}},\ \bibinfo {pages} {46}
  (\bibinfo {year} {2024})},\ \Eprint {http://arxiv.org/abs/2310.03231}
  {arXiv:2310.03231 [astro-ph.HE]} \BibitemShut {NoStop}%
\bibitem [{\citenamefont {Gallant}\ and\ \citenamefont
  {Achterberg}(1999)}]{Gallant:1998uq}%
  \BibitemOpen
  \bibfield  {author} {\bibinfo {author} {\bibfnamefont {Y.~A.}\ \bibnamefont
  {Gallant}}\ and\ \bibinfo {author} {\bibfnamefont {A.}~\bibnamefont
  {Achterberg}},\ }\href {\doibase 10.1046/j.1365-8711.1999.02566.x} {\bibfield
   {journal} {\bibinfo  {journal} {Mon. Not. Roy. Astron. Soc.}\ }\textbf
  {\bibinfo {volume} {305}},\ \bibinfo {pages} {6} (\bibinfo {year} {1999})},\
  \Eprint {http://arxiv.org/abs/astro-ph/9812316} {arXiv:astro-ph/9812316
  [astro-ph]} \BibitemShut {NoStop}%
\bibitem [{\citenamefont {Wang}\ \emph {et~al.}(2024)\citenamefont {Wang},
  \citenamefont {Reville}, \citenamefont {Rieger},\ and\ \citenamefont
  {Aharonian}}]{Wang:2024ijr}%
  \BibitemOpen
  \bibfield  {author} {\bibinfo {author} {\bibfnamefont {J.-S.}\ \bibnamefont
  {Wang}}, \bibinfo {author} {\bibfnamefont {B.}~\bibnamefont {Reville}},
  \bibinfo {author} {\bibfnamefont {F.~M.}\ \bibnamefont {Rieger}}, \ and\
  \bibinfo {author} {\bibfnamefont {F.~A.}\ \bibnamefont {Aharonian}},\ }\href
  {\doibase 10.3847/2041-8213/ad9589} {\bibfield  {journal} {\bibinfo
  {journal} {Astrophys. J. Lett.}\ }\textbf {\bibinfo {volume} {977}},\
  \bibinfo {pages} {L20} (\bibinfo {year} {2024})},\ \Eprint
  {http://arxiv.org/abs/2411.16674} {arXiv:2411.16674 [astro-ph.HE]}
  \BibitemShut {NoStop}%
\bibitem [{\citenamefont {{Grebenyuk}}\ \emph {et~al.}(2019)\citenamefont
  {{Grebenyuk}}, \citenamefont {{Karmanov}}, \citenamefont {{Kovalev}},
  \citenamefont {{Kudryashov}}, \citenamefont {{Kurganov}}, \citenamefont
  {{Panov}}, \citenamefont {{Podorozhny}}, \citenamefont {{Tkachenko}},
  \citenamefont {{Tkachev}}, \citenamefont {{Turundaevskiy}}, \citenamefont
  {{Vasiliev}},\ and\ \citenamefont {{Voronin}}}]{2019AdSpR..64.2546G}%
  \BibitemOpen
  \bibfield  {author} {\bibinfo {author} {\bibfnamefont {V.}~\bibnamefont
  {{Grebenyuk}}}, \bibinfo {author} {\bibfnamefont {D.}~\bibnamefont
  {{Karmanov}}}, \bibinfo {author} {\bibfnamefont {I.}~\bibnamefont
  {{Kovalev}}}, \bibinfo {author} {\bibfnamefont {I.}~\bibnamefont
  {{Kudryashov}}}, \bibinfo {author} {\bibfnamefont {A.}~\bibnamefont
  {{Kurganov}}}, \bibinfo {author} {\bibfnamefont {A.}~\bibnamefont {{Panov}}},
  \bibinfo {author} {\bibfnamefont {D.}~\bibnamefont {{Podorozhny}}}, \bibinfo
  {author} {\bibfnamefont {A.}~\bibnamefont {{Tkachenko}}}, \bibinfo {author}
  {\bibfnamefont {L.}~\bibnamefont {{Tkachev}}}, \bibinfo {author}
  {\bibfnamefont {A.}~\bibnamefont {{Turundaevskiy}}}, \bibinfo {author}
  {\bibfnamefont {O.}~\bibnamefont {{Vasiliev}}}, \ and\ \bibinfo {author}
  {\bibfnamefont {A.}~\bibnamefont {{Voronin}}},\ }\href {\doibase
  10.1016/j.asr.2019.10.004} {\bibfield  {journal} {\bibinfo  {journal}
  {Advances in Space Research}\ }\textbf {\bibinfo {volume} {64}},\ \bibinfo
  {pages} {2546} (\bibinfo {year} {2019})}\BibitemShut {NoStop}%
\bibitem [{\citenamefont {Matthews}(2017)}]{Matthews:2017rW}%
  \BibitemOpen
  \bibfield  {author} {\bibinfo {author} {\bibfnamefont {J.}~\bibnamefont
  {Matthews}},\ }\href {\doibase 10.22323/1.301.1096} {\bibfield  {journal}
  {\bibinfo  {journal} {PoS}\ }\textbf {\bibinfo {volume} {ICRC2017}},\
  \bibinfo {pages} {1096} (\bibinfo {year} {2017})}\BibitemShut {NoStop}%
\bibitem [{\citenamefont {Verzi}(2019)}]{Verzi:2019AO}%
  \BibitemOpen
  \bibfield  {author} {\bibinfo {author} {\bibfnamefont {V.}~\bibnamefont
  {Verzi}},\ }\href {\doibase 10.22323/1.358.0450} {\bibfield  {journal}
  {\bibinfo  {journal} {PoS}\ }\textbf {\bibinfo {volume} {ICRC2019}},\
  \bibinfo {pages} {450} (\bibinfo {year} {2019})}\BibitemShut {NoStop}%
\bibitem [{\citenamefont {Arteaga-Velázquez}\ \emph
  {et~al.}(2017)\citenamefont {Arteaga-Velázquez}, \citenamefont
  {Rivera-Rangel}, \citenamefont {Apel}, \citenamefont {Bekk}, \citenamefont
  {Bertaina}, \citenamefont {Blümer}, \citenamefont {Bozdog}, \citenamefont
  {Brancus}, \citenamefont {Cantoni}, \citenamefont {CHIAVASSA}, \citenamefont
  {Cossavella}, \citenamefont {Daumiller}, \citenamefont {de~Souza},
  \citenamefont {Di~Pierro}, \citenamefont {Doll}, \citenamefont {Engel},
  \citenamefont {Fuhrmann}, \citenamefont {Gherghel-Lascu}, \citenamefont
  {Gils}, \citenamefont {Glasstetter}, \citenamefont {Grupen}, \citenamefont
  {Haungs}, \citenamefont {Heck}, \citenamefont {Hörandel}, \citenamefont
  {Huber}, \citenamefont {Huege}, \citenamefont {Kampert}, \citenamefont
  {Kang}, \citenamefont {Klages}, \citenamefont {Link}, \citenamefont
  {Łuczak}, \citenamefont {Mathes}, \citenamefont {Mayer}, \citenamefont
  {Milke}, \citenamefont {Mitrica}, \citenamefont {Morello}, \citenamefont
  {Oehlschläger}, \citenamefont {Ostapchenko}, \citenamefont {Palmieri},
  \citenamefont {Pierog}, \citenamefont {Rebel}, \citenamefont {Roth},
  \citenamefont {Schieler}, \citenamefont {Schoo}, \citenamefont {Schröder},
  \citenamefont {Sima}, \citenamefont {Toma}, \citenamefont {Trinchero},
  \citenamefont {Ulrich}, \citenamefont {Weindl}, \citenamefont {Wochele},\
  and\ \citenamefont {Zabierowski}}]{Arteaga-Velázquez:2017/0}%
  \BibitemOpen
  \bibfield  {author} {\bibinfo {author} {\bibfnamefont {J.~C.}\ \bibnamefont
  {Arteaga-Velázquez}}, \bibinfo {author} {\bibfnamefont {D.}~\bibnamefont
  {Rivera-Rangel}}, \bibinfo {author} {\bibfnamefont {W.}~\bibnamefont {Apel}},
  \bibinfo {author} {\bibfnamefont {K.}~\bibnamefont {Bekk}}, \bibinfo {author}
  {\bibfnamefont {M.~E.}\ \bibnamefont {Bertaina}}, \bibinfo {author}
  {\bibfnamefont {J.}~\bibnamefont {Blümer}}, \bibinfo {author} {\bibfnamefont
  {H.}~\bibnamefont {Bozdog}}, \bibinfo {author} {\bibfnamefont
  {I.}~\bibnamefont {Brancus}}, \bibinfo {author} {\bibfnamefont
  {E.}~\bibnamefont {Cantoni}}, \bibinfo {author} {\bibfnamefont
  {E.}~\bibnamefont {CHIAVASSA}}, \bibinfo {author} {\bibfnamefont
  {F.}~\bibnamefont {Cossavella}}, \bibinfo {author} {\bibfnamefont
  {K.}~\bibnamefont {Daumiller}}, \bibinfo {author} {\bibfnamefont
  {V.}~\bibnamefont {de~Souza}}, \bibinfo {author} {\bibfnamefont
  {F.}~\bibnamefont {Di~Pierro}}, \bibinfo {author} {\bibfnamefont
  {P.}~\bibnamefont {Doll}}, \bibinfo {author} {\bibfnamefont {R.}~\bibnamefont
  {Engel}}, \bibinfo {author} {\bibfnamefont {D.}~\bibnamefont {Fuhrmann}},
  \bibinfo {author} {\bibfnamefont {A.}~\bibnamefont {Gherghel-Lascu}},
  \bibinfo {author} {\bibfnamefont {H.}~\bibnamefont {Gils}}, \bibinfo {author}
  {\bibfnamefont {R.}~\bibnamefont {Glasstetter}}, \bibinfo {author}
  {\bibfnamefont {C.}~\bibnamefont {Grupen}}, \bibinfo {author} {\bibfnamefont
  {A.}~\bibnamefont {Haungs}}, \bibinfo {author} {\bibfnamefont
  {D.}~\bibnamefont {Heck}}, \bibinfo {author} {\bibfnamefont {J.}~\bibnamefont
  {Hörandel}}, \bibinfo {author} {\bibfnamefont {D.}~\bibnamefont {Huber}},
  \bibinfo {author} {\bibfnamefont {T.}~\bibnamefont {Huege}}, \bibinfo
  {author} {\bibfnamefont {K.-H.}\ \bibnamefont {Kampert}}, \bibinfo {author}
  {\bibfnamefont {D.}~\bibnamefont {Kang}}, \bibinfo {author} {\bibfnamefont
  {H.}~\bibnamefont {Klages}}, \bibinfo {author} {\bibfnamefont
  {K.}~\bibnamefont {Link}}, \bibinfo {author} {\bibfnamefont {P.}~\bibnamefont
  {Łuczak}}, \bibinfo {author} {\bibfnamefont {H.}~\bibnamefont {Mathes}},
  \bibinfo {author} {\bibfnamefont {H.}~\bibnamefont {Mayer}}, \bibinfo
  {author} {\bibfnamefont {J.}~\bibnamefont {Milke}}, \bibinfo {author}
  {\bibfnamefont {B.}~\bibnamefont {Mitrica}}, \bibinfo {author} {\bibfnamefont
  {C.}~\bibnamefont {Morello}}, \bibinfo {author} {\bibfnamefont
  {J.}~\bibnamefont {Oehlschläger}}, \bibinfo {author} {\bibfnamefont
  {S.}~\bibnamefont {Ostapchenko}}, \bibinfo {author} {\bibfnamefont
  {N.}~\bibnamefont {Palmieri}}, \bibinfo {author} {\bibfnamefont
  {T.}~\bibnamefont {Pierog}}, \bibinfo {author} {\bibfnamefont
  {H.}~\bibnamefont {Rebel}}, \bibinfo {author} {\bibfnamefont
  {M.}~\bibnamefont {Roth}}, \bibinfo {author} {\bibfnamefont {H.}~\bibnamefont
  {Schieler}}, \bibinfo {author} {\bibfnamefont {S.}~\bibnamefont {Schoo}},
  \bibinfo {author} {\bibfnamefont {F.}~\bibnamefont {Schröder}}, \bibinfo
  {author} {\bibfnamefont {O.}~\bibnamefont {Sima}}, \bibinfo {author}
  {\bibfnamefont {G.}~\bibnamefont {Toma}}, \bibinfo {author} {\bibfnamefont
  {G.}~\bibnamefont {Trinchero}}, \bibinfo {author} {\bibfnamefont
  {H.}~\bibnamefont {Ulrich}}, \bibinfo {author} {\bibfnamefont
  {A.}~\bibnamefont {Weindl}}, \bibinfo {author} {\bibfnamefont
  {J.}~\bibnamefont {Wochele}}, \ and\ \bibinfo {author} {\bibfnamefont
  {J.}~\bibnamefont {Zabierowski}},\ }\href {\doibase 10.22323/1.301.0316}
  {\bibfield  {journal} {\bibinfo  {journal} {PoS}\ }\textbf {\bibinfo {volume}
  {ICRC2017}},\ \bibinfo {pages} {316} (\bibinfo {year} {2017})}\BibitemShut
  {NoStop}%
\bibitem [{\citenamefont {Aartsen}\ \emph {et~al.}(2019)\citenamefont {Aartsen}
  \emph {et~al.}}]{IceCube:2019hmk}%
  \BibitemOpen
  \bibfield  {author} {\bibinfo {author} {\bibfnamefont {M.~G.}\ \bibnamefont
  {Aartsen}} \emph {et~al.} (\bibinfo {collaboration} {IceCube}),\ }\href
  {\doibase 10.1103/PhysRevD.100.082002} {\bibfield  {journal} {\bibinfo
  {journal} {Phys. Rev. D}\ }\textbf {\bibinfo {volume} {100}},\ \bibinfo
  {pages} {082002} (\bibinfo {year} {2019})},\ \Eprint
  {http://arxiv.org/abs/1906.04317} {arXiv:1906.04317 [astro-ph.HE]}
  \BibitemShut {NoStop}%
\bibitem [{\citenamefont {{Parker}}(1965)}]{1965P&SS...13....9P}%
  \BibitemOpen
  \bibfield  {author} {\bibinfo {author} {\bibfnamefont {E.~N.}\ \bibnamefont
  {{Parker}}},\ }\href {\doibase 10.1016/0032-0633(65)90131-5} {\bibfield
  {journal} {\bibinfo  {journal} {Planet. Space Sci.}\ }\textbf {\bibinfo
  {volume} {13}},\ \bibinfo {pages} {9} (\bibinfo {year} {1965})}\BibitemShut
  {NoStop}%
\bibitem [{\citenamefont {Gaisser}\ \emph {et~al.}(2016)\citenamefont
  {Gaisser}, \citenamefont {Engel},\ and\ \citenamefont
  {Resconi}}]{Gaisser:2016uoy}%
  \BibitemOpen
  \bibfield  {author} {\bibinfo {author} {\bibfnamefont {T.~K.}\ \bibnamefont
  {Gaisser}}, \bibinfo {author} {\bibfnamefont {R.}~\bibnamefont {Engel}}, \
  and\ \bibinfo {author} {\bibfnamefont {E.}~\bibnamefont {Resconi}},\
  }\href@noop {} {\emph {\bibinfo {title} {{Cosmic Rays and Particle Physics}:
  {2nd Edition}}}}\ (\bibinfo  {publisher} {Cambridge University Press},\
  \bibinfo {year} {2016})\BibitemShut {NoStop}%
\bibitem [{\citenamefont {Evoli}\ \emph {et~al.}(2017)\citenamefont {Evoli},
  \citenamefont {Gaggero}, \citenamefont {Vittino}, \citenamefont
  {Di~Bernardo}, \citenamefont {Di~Mauro}, \citenamefont {Ligorini},
  \citenamefont {Ullio},\ and\ \citenamefont {Grasso}}]{Evoli:2016xgn}%
  \BibitemOpen
  \bibfield  {author} {\bibinfo {author} {\bibfnamefont {C.}~\bibnamefont
  {Evoli}}, \bibinfo {author} {\bibfnamefont {D.}~\bibnamefont {Gaggero}},
  \bibinfo {author} {\bibfnamefont {A.}~\bibnamefont {Vittino}}, \bibinfo
  {author} {\bibfnamefont {G.}~\bibnamefont {Di~Bernardo}}, \bibinfo {author}
  {\bibfnamefont {M.}~\bibnamefont {Di~Mauro}}, \bibinfo {author}
  {\bibfnamefont {A.}~\bibnamefont {Ligorini}}, \bibinfo {author}
  {\bibfnamefont {P.}~\bibnamefont {Ullio}}, \ and\ \bibinfo {author}
  {\bibfnamefont {D.}~\bibnamefont {Grasso}},\ }\href {\doibase
  10.1088/1475-7516/2017/02/015} {\bibfield  {journal} {\bibinfo  {journal}
  {JCAP}\ }\textbf {\bibinfo {volume} {02}},\ \bibinfo {pages} {015} (\bibinfo
  {year} {2017})},\ \Eprint {http://arxiv.org/abs/1607.07886} {arXiv:1607.07886
  [astro-ph.HE]} \BibitemShut {NoStop}%
\bibitem [{\citenamefont {Ferriere}(2001)}]{Ferriere:2001rg}%
  \BibitemOpen
  \bibfield  {author} {\bibinfo {author} {\bibfnamefont {K.~M.}\ \bibnamefont
  {Ferriere}},\ }\href {\doibase 10.1103/RevModPhys.73.1031} {\bibfield
  {journal} {\bibinfo  {journal} {Rev. Mod. Phys.}\ }\textbf {\bibinfo {volume}
  {73}},\ \bibinfo {pages} {1031} (\bibinfo {year} {2001})},\ \Eprint
  {http://arxiv.org/abs/astro-ph/0106359} {arXiv:astro-ph/0106359} \BibitemShut
  {NoStop}%
\bibitem [{\citenamefont {Murase}\ and\ \citenamefont
  {Fukugita}(2019)}]{Murase:2018utn}%
  \BibitemOpen
  \bibfield  {author} {\bibinfo {author} {\bibfnamefont {K.}~\bibnamefont
  {Murase}}\ and\ \bibinfo {author} {\bibfnamefont {M.}~\bibnamefont
  {Fukugita}},\ }\href {\doibase 10.1103/PhysRevD.99.063012} {\bibfield
  {journal} {\bibinfo  {journal} {Phys. Rev. D}\ }\textbf {\bibinfo {volume}
  {99}},\ \bibinfo {pages} {063012} (\bibinfo {year} {2019})},\ \Eprint
  {http://arxiv.org/abs/1806.04194} {arXiv:1806.04194 [astro-ph.HE]}
  \BibitemShut {NoStop}%
\bibitem [{\citenamefont {Kimura}\ \emph
  {et~al.}(2018{\natexlab{b}})\citenamefont {Kimura}, \citenamefont {Murase},\
  and\ \citenamefont {M\'esz\'aros}}]{Kimura:2018ggg}%
  \BibitemOpen
  \bibfield  {author} {\bibinfo {author} {\bibfnamefont {S.~S.}\ \bibnamefont
  {Kimura}}, \bibinfo {author} {\bibfnamefont {K.}~\bibnamefont {Murase}}, \
  and\ \bibinfo {author} {\bibfnamefont {P.}~\bibnamefont {M\'esz\'aros}},\
  }\href {\doibase 10.3847/1538-4357/aadc0a} {\bibfield  {journal} {\bibinfo
  {journal} {Astrophys. J.}\ }\textbf {\bibinfo {volume} {866}},\ \bibinfo
  {pages} {51} (\bibinfo {year} {2018}{\natexlab{b}})},\ \Eprint
  {http://arxiv.org/abs/1807.03290} {arXiv:1807.03290 [astro-ph.HE]}
  \BibitemShut {NoStop}%
\bibitem [{\citenamefont {Aguilar}\ \emph {et~al.}(2016)\citenamefont {Aguilar}
  \emph {et~al.}}]{AMS:2016brs}%
  \BibitemOpen
  \bibfield  {author} {\bibinfo {author} {\bibfnamefont {M.}~\bibnamefont
  {Aguilar}} \emph {et~al.} (\bibinfo {collaboration} {AMS}),\ }\href {\doibase
  10.1103/PhysRevLett.117.231102} {\bibfield  {journal} {\bibinfo  {journal}
  {Phys. Rev. Lett.}\ }\textbf {\bibinfo {volume} {117}},\ \bibinfo {pages}
  {231102} (\bibinfo {year} {2016})}\BibitemShut {NoStop}%
\bibitem [{\citenamefont {Aguilar}\ \emph
  {et~al.}(2018{\natexlab{a}})\citenamefont {Aguilar} \emph
  {et~al.}}]{AMS:2018qxq}%
  \BibitemOpen
  \bibfield  {author} {\bibinfo {author} {\bibfnamefont {M.}~\bibnamefont
  {Aguilar}} \emph {et~al.} (\bibinfo {collaboration} {AMS}),\ }\href {\doibase
  10.1103/PhysRevLett.121.051101} {\bibfield  {journal} {\bibinfo  {journal}
  {Phys. Rev. Lett.}\ }\textbf {\bibinfo {volume} {121}},\ \bibinfo {pages}
  {051101} (\bibinfo {year} {2018}{\natexlab{a}})}\BibitemShut {NoStop}%
\bibitem [{\citenamefont {Aguilar}\ \emph
  {et~al.}(2018{\natexlab{b}})\citenamefont {Aguilar} \emph
  {et~al.}}]{AMS:2018cen}%
  \BibitemOpen
  \bibfield  {author} {\bibinfo {author} {\bibfnamefont {M.}~\bibnamefont
  {Aguilar}} \emph {et~al.} (\bibinfo {collaboration} {AMS}),\ }\href {\doibase
  10.1103/PhysRevLett.121.051103} {\bibfield  {journal} {\bibinfo  {journal}
  {Phys. Rev. Lett.}\ }\textbf {\bibinfo {volume} {121}},\ \bibinfo {pages}
  {051103} (\bibinfo {year} {2018}{\natexlab{b}})}\BibitemShut {NoStop}%
\bibitem [{\citenamefont {Aguilar}\ \emph {et~al.}(2015)\citenamefont {Aguilar}
  \emph {et~al.}}]{AMS:2015tnn}%
  \BibitemOpen
  \bibfield  {author} {\bibinfo {author} {\bibfnamefont {M.}~\bibnamefont
  {Aguilar}} \emph {et~al.} (\bibinfo {collaboration} {AMS}),\ }\href {\doibase
  10.1103/PhysRevLett.114.171103} {\bibfield  {journal} {\bibinfo  {journal}
  {Phys. Rev. Lett.}\ }\textbf {\bibinfo {volume} {114}},\ \bibinfo {pages}
  {171103} (\bibinfo {year} {2015})}\BibitemShut {NoStop}%
\bibitem [{\citenamefont {Aguilar}\ \emph {et~al.}(2021)\citenamefont {Aguilar}
  \emph {et~al.}}]{AMS:2021lxc}%
  \BibitemOpen
  \bibfield  {author} {\bibinfo {author} {\bibfnamefont {M.}~\bibnamefont
  {Aguilar}} \emph {et~al.} (\bibinfo {collaboration} {AMS}),\ }\href {\doibase
  10.1103/PhysRevLett.126.041104} {\bibfield  {journal} {\bibinfo  {journal}
  {Phys. Rev. Lett.}\ }\textbf {\bibinfo {volume} {126}},\ \bibinfo {pages}
  {041104} (\bibinfo {year} {2021})}\BibitemShut {NoStop}%
\bibitem [{\citenamefont {Aguilar}\ \emph {et~al.}(2020)\citenamefont {Aguilar}
  \emph {et~al.}}]{AMS:2020cai}%
  \BibitemOpen
  \bibfield  {author} {\bibinfo {author} {\bibfnamefont {M.}~\bibnamefont
  {Aguilar}} \emph {et~al.} (\bibinfo {collaboration} {AMS}),\ }\href {\doibase
  10.1103/PhysRevLett.124.211102} {\bibfield  {journal} {\bibinfo  {journal}
  {Phys. Rev. Lett.}\ }\textbf {\bibinfo {volume} {124}},\ \bibinfo {pages}
  {211102} (\bibinfo {year} {2020})}\BibitemShut {NoStop}%
\bibitem [{\citenamefont {Lodders}(2010)}]{Lodders:2010kx}%
  \BibitemOpen
  \bibfield  {author} {\bibinfo {author} {\bibfnamefont {K.}~\bibnamefont
  {Lodders}}\ }(\bibinfo {year} {2010})\ pp.\ \bibinfo {pages} {379--417},\
  \Eprint {http://arxiv.org/abs/1010.2746} {arXiv:1010.2746 [astro-ph.SR]}
  \BibitemShut {NoStop}%
\bibitem [{\citenamefont {Caprioli}\ \emph {et~al.}(2017)\citenamefont
  {Caprioli}, \citenamefont {Yi},\ and\ \citenamefont
  {Spitkovsky}}]{Caprioli:2017oun}%
  \BibitemOpen
  \bibfield  {author} {\bibinfo {author} {\bibfnamefont {D.}~\bibnamefont
  {Caprioli}}, \bibinfo {author} {\bibfnamefont {D.~T.}\ \bibnamefont {Yi}}, \
  and\ \bibinfo {author} {\bibfnamefont {A.}~\bibnamefont {Spitkovsky}},\
  }\href {\doibase 10.1103/PhysRevLett.119.171101} {\bibfield  {journal}
  {\bibinfo  {journal} {Phys. Rev. Lett.}\ }\textbf {\bibinfo {volume} {119}},\
  \bibinfo {pages} {171101} (\bibinfo {year} {2017})},\ \Eprint
  {http://arxiv.org/abs/1704.08252} {arXiv:1704.08252 [astro-ph.HE]}
  \BibitemShut {NoStop}%
\bibitem [{\citenamefont {Murase}\ \emph {et~al.}(2008)\citenamefont {Murase},
  \citenamefont {Inoue},\ and\ \citenamefont {Nagataki}}]{Murase:2008yt}%
  \BibitemOpen
  \bibfield  {author} {\bibinfo {author} {\bibfnamefont {K.}~\bibnamefont
  {Murase}}, \bibinfo {author} {\bibfnamefont {S.}~\bibnamefont {Inoue}}, \
  and\ \bibinfo {author} {\bibfnamefont {S.}~\bibnamefont {Nagataki}},\ }\href
  {\doibase 10.1086/595882} {\bibfield  {journal} {\bibinfo  {journal}
  {Astrophys. J.}\ }\textbf {\bibinfo {volume} {689}},\ \bibinfo {pages} {L105}
  (\bibinfo {year} {2008})},\ \Eprint {http://arxiv.org/abs/0805.0104}
  {arXiv:0805.0104 [astro-ph]} \BibitemShut {NoStop}%
\bibitem [{\citenamefont {Yushkov}(2020)}]{Yushkov:2020nhr}%
  \BibitemOpen
  \bibfield  {author} {\bibinfo {author} {\bibfnamefont {A.}~\bibnamefont
  {Yushkov}} (\bibinfo {collaboration} {Auger}),\ }\href {\doibase
  10.22323/1.358.0482} {\bibfield  {journal} {\bibinfo  {journal} {PoS}\
  }\textbf {\bibinfo {volume} {ICRC2019}},\ \bibinfo {pages} {482} (\bibinfo
  {year} {2020})}\BibitemShut {NoStop}%
\bibitem [{\citenamefont {Apel}\ \emph {et~al.}(2013)\citenamefont {Apel} \emph
  {et~al.}}]{Apel:2013uni}%
  \BibitemOpen
  \bibfield  {author} {\bibinfo {author} {\bibfnamefont {W.~D.}\ \bibnamefont
  {Apel}} \emph {et~al.},\ }\href {\doibase
  10.1016/j.astropartphys.2013.06.004} {\bibfield  {journal} {\bibinfo
  {journal} {Astropart. Phys.}\ }\textbf {\bibinfo {volume} {47}},\ \bibinfo
  {pages} {54} (\bibinfo {year} {2013})},\ \Eprint
  {http://arxiv.org/abs/1306.6283} {arXiv:1306.6283 [astro-ph.HE]} \BibitemShut
  {NoStop}%
\bibitem [{\citenamefont {Murase}\ \emph {et~al.}(2013)\citenamefont {Murase},
  \citenamefont {Ahlers},\ and\ \citenamefont {Lacki}}]{Murase:2013rfa}%
  \BibitemOpen
  \bibfield  {author} {\bibinfo {author} {\bibfnamefont {K.}~\bibnamefont
  {Murase}}, \bibinfo {author} {\bibfnamefont {M.}~\bibnamefont {Ahlers}}, \
  and\ \bibinfo {author} {\bibfnamefont {B.~C.}\ \bibnamefont {Lacki}},\ }\href
  {\doibase 10.1103/PhysRevD.88.121301} {\bibfield  {journal} {\bibinfo
  {journal} {Phys. Rev. D}\ }\textbf {\bibinfo {volume} {88}},\ \bibinfo
  {pages} {121301} (\bibinfo {year} {2013})},\ \Eprint
  {http://arxiv.org/abs/1306.3417} {arXiv:1306.3417 [astro-ph.HE]} \BibitemShut
  {NoStop}%
\bibitem [{\citenamefont {Simeon}\ \emph {et~al.}(2025)\citenamefont {Simeon},
  \citenamefont {Globus}, \citenamefont {Barrow},\ and\ \citenamefont
  {Blandford}}]{Simeon:2025gxd}%
  \BibitemOpen
  \bibfield  {author} {\bibinfo {author} {\bibfnamefont {P.}~\bibnamefont
  {Simeon}}, \bibinfo {author} {\bibfnamefont {N.}~\bibnamefont {Globus}},
  \bibinfo {author} {\bibfnamefont {K.~S.~S.}\ \bibnamefont {Barrow}}, \ and\
  \bibinfo {author} {\bibfnamefont {R.}~\bibnamefont {Blandford}},\ }\href@noop
  {} {\  (\bibinfo {year} {2025})},\ \Eprint {http://arxiv.org/abs/2503.10795}
  {arXiv:2503.10795 [astro-ph.HE]} \BibitemShut {NoStop}%
\bibitem [{\citenamefont {Thoudam}\ \emph {et~al.}(2016)\citenamefont
  {Thoudam}, \citenamefont {Rachen}, \citenamefont {van Vliet}, \citenamefont
  {Achterberg}, \citenamefont {Buitink}, \citenamefont {Falcke},\ and\
  \citenamefont {H\"orandel}}]{Thoudam:2016syr}%
  \BibitemOpen
  \bibfield  {author} {\bibinfo {author} {\bibfnamefont {S.}~\bibnamefont
  {Thoudam}}, \bibinfo {author} {\bibfnamefont {J.~P.}\ \bibnamefont {Rachen}},
  \bibinfo {author} {\bibfnamefont {A.}~\bibnamefont {van Vliet}}, \bibinfo
  {author} {\bibfnamefont {A.}~\bibnamefont {Achterberg}}, \bibinfo {author}
  {\bibfnamefont {S.}~\bibnamefont {Buitink}}, \bibinfo {author} {\bibfnamefont
  {H.}~\bibnamefont {Falcke}}, \ and\ \bibinfo {author} {\bibfnamefont {J.~R.}\
  \bibnamefont {H\"orandel}},\ }\href {\doibase 10.1051/0004-6361/201628894}
  {\bibfield  {journal} {\bibinfo  {journal} {Astron. Astrophys.}\ }\textbf
  {\bibinfo {volume} {595}},\ \bibinfo {pages} {A33} (\bibinfo {year}
  {2016})},\ \Eprint {http://arxiv.org/abs/1605.03111} {arXiv:1605.03111
  [astro-ph.HE]} \BibitemShut {NoStop}%
\bibitem [{\citenamefont {Mukhopadhyay}\ \emph {et~al.}(2023)\citenamefont
  {Mukhopadhyay}, \citenamefont {Peretti}, \citenamefont {Globus},
  \citenamefont {Simeon},\ and\ \citenamefont
  {Blandford}}]{Mukhopadhyay:2023kel}%
  \BibitemOpen
  \bibfield  {author} {\bibinfo {author} {\bibfnamefont {P.}~\bibnamefont
  {Mukhopadhyay}}, \bibinfo {author} {\bibfnamefont {E.}~\bibnamefont
  {Peretti}}, \bibinfo {author} {\bibfnamefont {N.}~\bibnamefont {Globus}},
  \bibinfo {author} {\bibfnamefont {P.}~\bibnamefont {Simeon}}, \ and\ \bibinfo
  {author} {\bibfnamefont {R.}~\bibnamefont {Blandford}},\ }\href {\doibase
  10.3847/1538-4357/acdc9b} {\bibfield  {journal} {\bibinfo  {journal}
  {Astrophys. J.}\ }\textbf {\bibinfo {volume} {953}},\ \bibinfo {pages} {49}
  (\bibinfo {year} {2023})},\ \Eprint {http://arxiv.org/abs/2301.08902}
  {arXiv:2301.08902 [astro-ph.HE]} \BibitemShut {NoStop}%
\bibitem [{\citenamefont {Ptuskin}\ and\ \citenamefont
  {Zirakashvili}(2003)}]{Ptuskin:2003zv}%
  \BibitemOpen
  \bibfield  {author} {\bibinfo {author} {\bibfnamefont {V.~S.}\ \bibnamefont
  {Ptuskin}}\ and\ \bibinfo {author} {\bibfnamefont {V.~N.}\ \bibnamefont
  {Zirakashvili}},\ }\href {\doibase 10.1051/0004-6361:20030323} {\bibfield
  {journal} {\bibinfo  {journal} {Astron. Astrophys.}\ }\textbf {\bibinfo
  {volume} {403}},\ \bibinfo {pages} {1} (\bibinfo {year} {2003})},\ \Eprint
  {http://arxiv.org/abs/astro-ph/0302053} {arXiv:astro-ph/0302053} \BibitemShut
  {NoStop}%
\bibitem [{\citenamefont {{Ptuskin}}\ and\ \citenamefont
  {{Zirakashvili}}(2005)}]{2005A&A...429..755P}%
  \BibitemOpen
  \bibfield  {author} {\bibinfo {author} {\bibfnamefont {V.~S.}\ \bibnamefont
  {{Ptuskin}}}\ and\ \bibinfo {author} {\bibfnamefont {V.~N.}\ \bibnamefont
  {{Zirakashvili}}},\ }\href {\doibase 10.1051/0004-6361:20041517} {\bibfield
  {journal} {\bibinfo  {journal} {Astron.Astrophys.}\ }\textbf {\bibinfo
  {volume} {429}},\ \bibinfo {pages} {755} (\bibinfo {year} {2005})},\ \Eprint
  {http://arxiv.org/abs/astro-ph/0408025} {arXiv:astro-ph/0408025 [astro-ph]}
  \BibitemShut {NoStop}%
\bibitem [{\citenamefont {Kotera}\ \emph {et~al.}(2009)\citenamefont {Kotera},
  \citenamefont {Allard}, \citenamefont {Murase}, \citenamefont {Aoi},
  \citenamefont {Dubois}, \citenamefont {Pierog},\ and\ \citenamefont
  {Nagataki}}]{Kotera:2009ms}%
  \BibitemOpen
  \bibfield  {author} {\bibinfo {author} {\bibfnamefont {K.}~\bibnamefont
  {Kotera}}, \bibinfo {author} {\bibfnamefont {D.}~\bibnamefont {Allard}},
  \bibinfo {author} {\bibfnamefont {K.}~\bibnamefont {Murase}}, \bibinfo
  {author} {\bibfnamefont {J.}~\bibnamefont {Aoi}}, \bibinfo {author}
  {\bibfnamefont {Y.}~\bibnamefont {Dubois}}, \bibinfo {author} {\bibfnamefont
  {T.}~\bibnamefont {Pierog}}, \ and\ \bibinfo {author} {\bibfnamefont
  {S.}~\bibnamefont {Nagataki}},\ }\href {\doibase 10.1088/0004-637X/707/1/370}
  {\bibfield  {journal} {\bibinfo  {journal} {Astrophys. J.}\ }\textbf
  {\bibinfo {volume} {707}},\ \bibinfo {pages} {370} (\bibinfo {year}
  {2009})},\ \Eprint {http://arxiv.org/abs/0907.2433} {arXiv:0907.2433
  [astro-ph.HE]} \BibitemShut {NoStop}%
\bibitem [{\citenamefont {Fang}\ and\ \citenamefont
  {Murase}(2018)}]{Fang:2017zjf}%
  \BibitemOpen
  \bibfield  {author} {\bibinfo {author} {\bibfnamefont {K.}~\bibnamefont
  {Fang}}\ and\ \bibinfo {author} {\bibfnamefont {K.}~\bibnamefont {Murase}},\
  }\href {\doibase 10.1038/s41567-017-0025-4} {\bibfield  {journal} {\bibinfo
  {journal} {Nature Phys.}\ }\textbf {\bibinfo {volume} {14}},\ \bibinfo
  {pages} {396} (\bibinfo {year} {2018})},\ \Eprint
  {http://arxiv.org/abs/1704.00015} {arXiv:1704.00015 [astro-ph.HE]}
  \BibitemShut {NoStop}%
\bibitem [{\citenamefont {Gonz\'alez}\ \emph {et~al.}(2021)\citenamefont
  {Gonz\'alez}, \citenamefont {Mollerach},\ and\ \citenamefont
  {Roulet}}]{Gonzalez:2021ajv}%
  \BibitemOpen
  \bibfield  {author} {\bibinfo {author} {\bibfnamefont {J.~M.}\ \bibnamefont
  {Gonz\'alez}}, \bibinfo {author} {\bibfnamefont {S.}~\bibnamefont
  {Mollerach}}, \ and\ \bibinfo {author} {\bibfnamefont {E.}~\bibnamefont
  {Roulet}},\ }\href {\doibase 10.1103/PhysRevD.104.063005} {\bibfield
  {journal} {\bibinfo  {journal} {Phys. Rev. D}\ }\textbf {\bibinfo {volume}
  {104}},\ \bibinfo {pages} {063005} (\bibinfo {year} {2021})},\ \Eprint
  {http://arxiv.org/abs/2105.08138} {arXiv:2105.08138 [astro-ph.HE]}
  \BibitemShut {NoStop}%
\bibitem [{\citenamefont {Yamamoto}\ \emph {et~al.}(2008)\citenamefont
  {Yamamoto} \emph {et~al.}}]{Yamamoto:2008xr}%
  \BibitemOpen
  \bibfield  {author} {\bibinfo {author} {\bibfnamefont {H.}~\bibnamefont
  {Yamamoto}} \emph {et~al.},\ }\href {\doibase 10.1093/pasj/60.4.715}
  {\bibfield  {journal} {\bibinfo  {journal} {Publ. Astron. Soc. Jap.}\
  }\textbf {\bibinfo {volume} {60}},\ \bibinfo {pages} {715} (\bibinfo {year}
  {2008})},\ \Eprint {http://arxiv.org/abs/0804.1871} {arXiv:0804.1871
  [astro-ph]} \BibitemShut {NoStop}%
\bibitem [{\citenamefont {Giacinti}\ and\ \citenamefont
  {Semikoz}(2023)}]{Giacinti:2023upw}%
  \BibitemOpen
  \bibfield  {author} {\bibinfo {author} {\bibfnamefont {G.}~\bibnamefont
  {Giacinti}}\ and\ \bibinfo {author} {\bibfnamefont {D.}~\bibnamefont
  {Semikoz}},\ }\href@noop {} {\  (\bibinfo {year} {2023})},\ \Eprint
  {http://arxiv.org/abs/2305.10251} {arXiv:2305.10251 [astro-ph.HE]}
  \BibitemShut {NoStop}%
\bibitem [{\citenamefont {Stall}\ \emph {et~al.}(2025)\citenamefont {Stall},
  \citenamefont {Loo},\ and\ \citenamefont {Mertsch}}]{Stall:2024klf}%
  \BibitemOpen
  \bibfield  {author} {\bibinfo {author} {\bibfnamefont {A.}~\bibnamefont
  {Stall}}, \bibinfo {author} {\bibfnamefont {C.~K.}\ \bibnamefont {Loo}}, \
  and\ \bibinfo {author} {\bibfnamefont {P.}~\bibnamefont {Mertsch}},\ }\href
  {\doibase 10.3847/2041-8213/adaea8} {\bibfield  {journal} {\bibinfo
  {journal} {Astrophys. J. Lett.}\ }\textbf {\bibinfo {volume} {980}},\
  \bibinfo {pages} {L21} (\bibinfo {year} {2025})},\ \Eprint
  {http://arxiv.org/abs/2409.11012} {arXiv:2409.11012 [astro-ph.HE]}
  \BibitemShut {NoStop}%
\end{thebibliography}%

\end{document}